\documentclass[5p]{elsarticle}

\usepackage{lineno,hyperref}
\modulolinenumbers[5]

\usepackage[symbol]{footmisc}

\usepackage{float}
\usepackage{graphicx}
\usepackage{verbatim}
\usepackage{amsmath}
\usepackage{algorithm}
\usepackage{algpseudocode}
\usepackage{algorithmicx}
\usepackage{tabularx}
\usepackage{multirow}
\usepackage{hhline}
\usepackage{color}
\usepackage[capitalise]{cleveref}
\usepackage{float}
\usepackage{picins}

\crefname{section}{Sec.}{Secs.}

\usepackage{wrapfig}

\usepackage[nolist]{acronym}
\begin{acronym}
	\acro{AAA}{Authentication, Authorization and Accounting}
	\acro{AP}{Access Point}
	\acro{API}{Application Programming Interface}
	\acro{BP}{Baseline Pseudonym}
	\acro{BS}{Base Station} 
    \acro{BSM}{Basic Safety Message}
    \acro{BYOD}{Bring Your Own Device}
	\acro{C2C-CC}{Car2Car Communication Consortium}
	\acro{CA}{Certification Authority}
	\acrodefplural{CA}{Certification Authorities}
	\acro{CN}{Common Name}
	\acro{CAM}{Cooperative Awareness Message}
	\acro{CIA}{Confidentiality, Integrity and Availability}
	\acro{CRL}{Certificate Revocation List}
	\acro{CDN}{Content Delivery Network}
	\acro{CSR}{Certificate Signing Requests}
	\acro{DAA}{Direct Anonymous Attestation}
	\acro{DDoS}{Distributed Denial of Service}
	\acro{DDH}{Decisional Diffie-Helman}
	\acro{DENM}{Decentralized Environmental Notification Message}
	\acro{DHT}{Distributed Hash Table}
	\acro{DoS}{Denial of Service}
	\acro{DPA}{Data Protection Agency}
	\acro{ECU}{Electronic Control Unit}
	\acro{ETSI}{European Telecommunications Standards Institute}
	\acro{ECDSA}{Elliptic Curve Digital Signature Algorithm}
	\acro{ECC}{Elliptic Curve Cryptography}
	\acro{EVITA}{E-safety Vehicle Intrusion protected Applications}
	\acro{FCFS}{First-Come First-Served}
	\acro{FOT}{Field Operational Test}
	\acro{FPGA}{Field-Programmable Gate Array}
	\acro{GN}{GeoNetworking}
	\acro{GS}{Group Signatures}
	\acro{GM}{Group Manager}
	\acro{GBA}{Generic Bootstrapping Architecture}
	\acro{GPA}{Global Passive Adversary}
	\acro{GUI}{Graphic User Interface}
	\acro{HP}{Hybrid Pseudonym}
	\acro{HSM}{Hardware Security Module}
	\acro{HTTP}{Hypertext Transfer Protocol}
	\acro{IEEE}{Institute of Electrical and Electronics Engineers}
	\acro{IoT}{Internet of Things}
	\acro{ITS}{Intelligent Transport Systems}
	\acro{IT}{Information Technologies}
	\acro{IMSI}{International Mobile Subscriber Identity}
	\acro{IMEI}{International Mobile Station Equipment Identity}
	\acro{IdP}{Identity Provider}
	\acro{ISP}{Internet Service Provider}
	\acro{LCFS}{Last Come First Served} 
	\acro{LEA}{Law Enforcement Agency}
	\acro{LTC}{Long-Term Certificate}
	\acro{LTCA}{Long-Term \acl{CA}}
	\acrodefplural{LTCA}{Long-Term \aclp{CA}}
	\acro{LDAP}{Lightweight Directory Access Protocol}
	\acro{LTE}{Long Term Evolution}
	\acro{LBS}{Location-based Service}
	\acro{MAC}{Message Authentication Code}
	\acro{MCA}{Message \ac{CA}}
	\acro{MEA}{Misbehavior Evaluation Authority}
	\acro{OBU}{On-Board Unit}
	\acro{OCSP}{Online Certificate Status Protocol}
	\acro{OSN}{Online Social Network}
	\acro{PC}{Pseudonymous Certificate}
	\acro{PCA}{Pseudonymous \acl{CA}}
	\acrodefplural{PCA}{Pseudonymous \aclp{CA}}
	\acro{PDP}{Policy Decision Point}
	\acro{PEP}{Policy Enforcement Point}
	\acro{PIR}{Private Information Retrieval}
	
	\acro{PKI}{Public-Key Infrastructure}
	\acro{POI}{Point of Interest}
	\acro{PRECIOSA}{Privacy Enabled Capability in Co-operative Systems and Safety Applications}
	\acro{PRESERVE}{Preparing Secure Vehicle-to-X Communication Systems}
	\acro{P2P}{peer-to-peer}
	\acro{RA}{Resolution Authority}
	\acro{REST}{Representational State Transfer}
	\acro{RBAC}{Role Based Access Control}
	\acro{RCA}{Root \acl{CA}}
	\acro{RSU}{Roadside Unit}
	\acro{SAML}{Security Assertion Markup Language}
	\acro{SCORE@F}{Système COopératif Routier Expérimental Français}
	\acro{SeVeCom}{Secure Vehicle Communication}
	\acro{SIS}{Security Infrastructure Server}
	\acro{SP}{Service Provider}
	\acro{SSO}{Single-Sign-On}
	\acro{SoA}{Service-oriented-Approach}
	\acro{SOAP}{Simple Object Access Protocol}
	\acro{SAS}{Sample Aggregation Service}
	\acro{TESLA}{Timed Efficient Stream Loss-Tolerant Authentication}
	\acro{TS}{Task Service}
	\acro{TLS}{Transport Layer Security}
	\acro{TPM}{Trusted Platform Module}
	\acro{TTP}{Trusted Third Party}
	\acro{TVR}{Ticket Validation Repository}
	\acro{URI}{Uniform Resource Identifier}
	\acro{VANET}{Vehicular Ad-hoc Network}
	\acro{V2I}{Vehicle-to-Infrastructure}
	\acro{V2V}{Vehicle-to-Vehicle}
	\acro{V2X}{\ac{V2V} and/or \ac{V2I}}
	\acro{VC}{Vehicular Communication}
	\acro{VM}{Virtual Machine}
	\acro{VSN}{Vehicular Social Network}
	\acro{VSS}{\ac{VC} Security Subsystem}
	\acro{WAVE}{Wireless Access in Vehicular Environments}
	\acro{WSDL}{Web Services Discovery Language}
	\acro{W3C}{World Wide Web Consortium}
	\acro{VANET}{Vehicular Ad-hoc Network}
	\acro{VPKI}{Vehicular Public-Key Infrastructure}
	\acro{WS}{Web Service}
	\acro{WoT}{Web of Trust}
	\acro{WSACA}{\ac{WAVE} Service Advertisement \ac{CA}}
	\acro{XML}{Extensible Markup Language}
	\acro{XACML}{eXtensible Access Control Markup Language}
	\acro{3G}{3rd Generation}
\end{acronym}

\usepackage{algpseudocode}
\usepackage{varwidth}
\usepackage{subcaption}

\usepackage{algorithm}
\PassOptionsToPackage{noend}{algpseudocode}
\usepackage{algpseudocode}

\errorcontextlines\maxdimen

\makeatletter
\newcommand*{\algrule}[1][\algorithmicindent]{\makebox[#1][l]{\hspace*{.5em}\vrule height .75\baselineskip depth .25\baselineskip}}%

\newcount\ALG@printindent@tempcnta
\def\ALG@printindent{%
	\ifnum \theALG@nested>0
	\ifx\ALG@text\ALG@x@notext
	\addvspace{-3pt}
	\else
	\unskip
	\ALG@printindent@tempcnta=1
	\loop
	\algrule[\csname ALG@ind@\the\ALG@printindent@tempcnta\endcsname]%
	\advance \ALG@printindent@tempcnta 1
	\ifnum \ALG@printindent@tempcnta<\numexpr\theALG@nested+1\relax
	\repeat
	\fi
	\fi
}%
\usepackage{etoolbox}
\patchcmd{\ALG@doentity}{\noindent\hskip\ALG@tlm}{\ALG@printindent}{}{\errmessage{failed to patch}}
\makeatother

\definecolor{blue}{RGB}{0,0,0}

\begin{document}

\begin{frontmatter}

\title{DoS-resilient Cooperative Beacon Verification for Vehicular Communication Systems}

\author{Hongyu Jin\footnote[1]{}}
\ead{hongyuj@kth.se}
\cortext[cor1]{Corresponding author}

\author{Panos Papadimitratos}
\ead{papadim@kth.se}

\address{Networked Systems Security Group, KTH Royal Institute of Technology, Stockholm, Sweden}

\begin{abstract}
	Authenticated safety beacons in \ac{VC} systems ensure awareness among neighboring vehicles. However, the verification of beacon signatures introduces significant processing overhead for resource-constrained vehicular \acp{OBU}. Even worse in dense neighborhood or when a clogging \ac{DoS} attack is mounted. The \ac{OBU} would fail to verify for all received (authentic or fictitious) beacons. This could significantly delay the verifications of authentic beacons or even affect the awareness of neighboring vehicle status. In this paper, we propose an efficient cooperative beacon verification scheme leveraging efficient symmetric key based authentication on top of pseudonymous authentication (based on traditional public key cryptography), providing efficient discovery of authentic beacons among a pool of received authentic and fictitious beacons, and can significantly decrease waiting times of beacons in queue before their validations. We show with simulation results that our scheme can guarantee low waiting times for received beacons even in high neighbor density situations and under DoS attacks, under which a traditional scheme would not be workable.
\end{abstract}

\begin{keyword}
	Security\sep Privacy\sep Pseudonymous authentication\sep Efficiency
\end{keyword}

\end{frontmatter}

\section{Introduction}
\label{sec:introduction}

\ac{V2V} and \ac{V2I} communication in \acf{VC} systems entail high-rate transmissions: typically, for safety beacons, vehicles/\acfp{OBU} transmit at a rate of $10$ $Hz$, i.e., $10$ beacons per second. In spite of approaches that adapt the beacon rate, the challenge is clear: as \ac{VC} systems get progressively widely deployed, each vehicle will have to process safety beacons (along with other traffic) from several tens of vehicles within its \ac{OBU} range, e.g., 300 messages per second for 30 neighboring vehicles. The provision of security and privacy protection aggravates the situation, adding communication overhead (digital signatures and short-term certificates attached, thus longer messages), as well as computation overhead (signature verifications mostly, and signature calculations).

A number of improvements (e.g.,~\cite{calandriello2007efficient, calandriello2011performance,feiri2014formal}) are compatible with the standardized pseudonymous authentication approach. For example, the \acp{PC}, termed \emph{pseudonyms}, can be attached to beacons periodically or based on the sender's context~\cite{feiri2014formal}, instead of attaching on each and every beacon. At the receiver side, a \ac{PC} needs to be validated only once and cached~\cite{calandriello2011performance}. For newly received beacons attached with cached \acp{PC}, only the signatures on beacons need to be verified. These approaches provide significant improvements and show how one can dimension processing power~\cite{calandriello2007efficient, calandriello2011performance}, but they are conservative: they assume each node verifies signatures on all received beacons. Indeed, this is the straightforward approach. An alternative, adaptive, reactive approach has been considered in \cite{ristanovic2011adaptive}, but only for multi-hop messages.

In~\cite{jin2015scaling}, vehicles share beacon verification results, so that the vehicles can benefit from verification efforts of neighboring vehicles. However, the work only considered an environment in which only benign vehicles broadcast authentic beacons. {\color{blue} In fact, dynamic vehicular mobility and thus topology (connectivity and physical neighborhood) creates a simple yet very effective attack vector, a clogging \ac{DoS} attack: an attacker, even external, could generate large volumes of fictitious beacons, purportedly from not previously encountered vehicles/\acp{OBU}. This would essentially prevent timely reception and validation of legitimate beacons. Even more so if there are multiple such offending adversarial transmitters across an area of the \ac{VC} system - in which case, one could even classify this as a distributed attack, a \ac{DDoS} attack. Discovering valid pseudonyms is challenging, because each node receives a pool of valid and non-valid pseudonyms, while the non-valid pseudonyms could be the majority.} This attack is cheap, because the attacker only needs to generate random bytes as signatures and attaches them to the broadcasted masqueraded beacons, while all its neighbors would be affected. Although the shared verification results can expedite the validation process~\cite{jin2015scaling}, the beacon queue can be saturated by an extremely high fictitious beacon rate, inevitably resulting in significantly high waiting time for authentic beacons (the majority of computation resources has to be used to verify fictitious signatures). Moreover, a successful \ac{PC} verification does not guarantee the timely verifications of following beacons from the same sender, because adversaries can simply attach valid \acp{PC} to the fictitious beacons.


This is exactly the problem we address in this paper. We extend the design of cooperative beacon verification scheme proposed in~\cite{jin2015scaling} by taking into consideration the defense against \ac{DoS} attacks and leveraging efficient symmetric key based authentication. Here, we use \ac{TESLA}~\cite{perrig2000efficient} on top of (a traditional public key cryptography based) pseudonymous authentication. Vehicles cooperatively discover beacons signed under new authentic \acp{PC} and cache the \acp{PC}. The beacons attached with cached \acp{PC} can be validated based on (1) signature verifications, (2) shared verification results, or (3) TESLA \acp{MAC}. Our simulation results show our scheme can significantly decrease waiting time for authentic beacons in high neighbor density situations and under \ac{DDoS} attacks, where a traditional scheme would not be workable.

In the rest of the paper, we provide background information on \ac{VC} security and privacy and related works (\cref{sec:related}), we then explain the system and adversary models, and security requirements (\cref{sec:problem}). We present in detail our scheme (\cref{sec:scheme}), and provide a security analysis (\cref{sec:analysis}), followed by evaluation based on simulation results (\cref{sec:simulation}), before concluding remarks (\cref{sec:conclusion}).

\section{Background and related works}
\label{sec:related}

A basic functionality for \ac{VC} systems is safety beacons, used to inform vehicle status to surrounding/neighboring vehicles. This can in turn improve traffic efficiency and safety by virtue of the awareness of the transportation environment. Authentication and integrity of safety beacons are strictly required. Traditional public key cryptography could provide those, yet the use of a long-term key pair and certificate would undermine user privacy: safety beacons could be trivially used to continuously track the sender vehicles. Pseudonymous authentication~\cite{calandriello2011performance, gisdakis2013serosa, khodaei2014towards, khodaei2018secmace} provides both security and privacy for safety beacons. A \ac{PC} is a short-term certificate issued by the \ac{VPKI} without any information on the long-term identity of the vehicle, thus making messages signed under different \acp{PC} unlinkable. However, this anonymity/pseudonymity is conditional: when misbehavior is detected, the relevant \acp{PC} (corresponding to the misbehaving \ac{OBU}/vehicle) can be revealed through a resolution protocol and then revoked~\cite{calandriello2011performance, gisdakis2013serosa, khodaei2014towards, khodaei2018secmace}.

Signature verifications remain expensive for resource-constrained \acp{OBU}. Thus, optimizations for decreasing communication and computation overhead~\cite{calandriello2007efficient, calandriello2011performance} have been proposed, but they do not change the fact that the signatures on each and every received beacons should be verified. Cooperative beacon verification~\cite{jin2015scaling, lin2013achieving} can help validating beacons based on shared verification results, but it is not resilient to DoS attacks, because a large portion of computation resources would still be used to verify fictitious beacons. \acp{PC} can be validated based on a pre-downloaded bloom filter~\cite{jin2016proactive}, but \acp{PC} for newly joining vehicles (commonly so in \ac{VC} systems) after the generation (and downloading) of the bloom filter would not be included, thus signatures on those \acp{PC} need to be verified. This could still leave a gap for adversarial nodes to inject fictitious \acp{PC} to the network.

TESLA-based authentication for \ac{VC} systems~\cite{hu2006strong,studer2009flexible} provides a cheaper way for validating successive beacons by a given sender, after a successful signature verification. However, it is still critical to validate new \acp{PC} and the corresponding beacons in high node density environment or under \ac{DoS} attacks in a timely manner. Prediction of the next beacon content (e.g., vehicle location) can help validating beacons based on the prediction result included in a previous beacon~\cite{hsiao2011flooding, lyu2016pba}. However, such an approach cannot tolerate packet losses: once a beacon is lost, the following beacon can only be verified based on its signature. Integrating with TESLA-based authentication can address packet losses in the prediction-based approaches~\cite{lyu2016pba}. However, in a network with high packet loss ratio (in dense neighborhood or under \ac{DoS} attacks), the majority of beacons need to be validated based on TESLA MACs, while the prediction based authentication would be barely used.


Recall that vehicles need to change their \acp{PC} periodically and frequently. An adversary can target this feature and flood with fictitious beacons, which can significantly delay verifications (and affect the caching) of new \acp{PC}. This problem is not addressed by any of the works discussed above. Our scheme addresses this problem by providing a cooperative approach for efficiently discovering authentic \acp{PC} when a large amount of new (authentic and fictitious) \acp{PC} are received, and can provide timely beacon validation with shared verification results and TESLA.

{\color{blue}

Puzzle-based approaches~\cite{sun2017privacy,liu2018mitigating} have been proposed to defend against \ac{DoS} attacks for mutual authentication in \ac{V2I} and \ac{V2V} communication, but they are not suitable for authenticating frequent connectionless safety beacons. Efficient pseudonym validation schemes based on alternative non-classical cryptographic primitives (e.g., ID-based cryptography)~\cite{sulaiman2013improving,wasef2013emap,jiang2016efficient} rely on connection to \ac{RSU} and need to communicate with \acp{RSU} for updating pseudonyms of newly joined or revoked nodes, and could fail to validate pseudonyms when connection to \acp{RSU} is lost. An efficient Certificate Revocation List (CRL) release approach~\cite{khodaei2018efficient} is proposed to defend against DoS attacks on CRL checking, however, this is orthogonal to our scheme and the two can co-exist.

}
\section{System model and security requirements}
\label{sec:problem}

\subsection{System and adversary model}

\begin{figure}[h!]
	\centering
	\includegraphics[width=0.8\columnwidth]{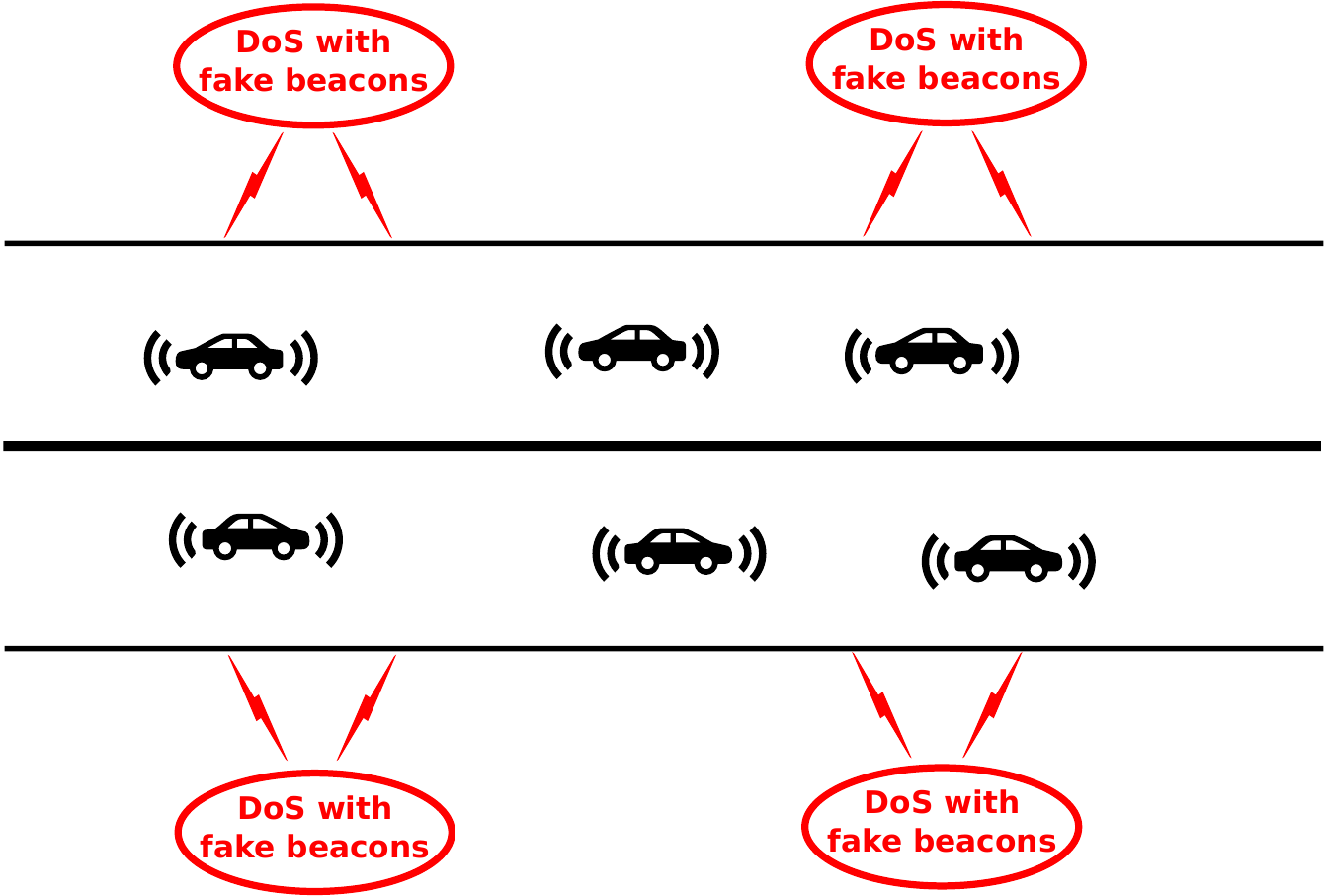}
	\caption{{\color{blue} Adversaries launching DDoS attacks.}}
	\label{fig_model}
\end{figure}

In our system, each vehicle is equipped with a set of \acp{PC} obtained from a \ac{VPKI}~\cite{khodaei2018secmace}. We assume a Sybil-resilient pseudonym lifetime policy as the one proposed in~\cite{khodaei2018secmace}: \acp{PC} have non-overlapping lifetimes and all pseudonym lifetimes are aligned. We mandate each and every beacon of a vehicle is properly signed with the private key corresponding to the currently valid \ac{PC}.

{\color{blue} We consider an overall landscape of adversarial models, and security and privacy requirements~\cite{papadimitratos2006securing} for our scheme design, while, in this paper, we are primarily concerned with adversaries that seek to exhaust computation resources of \acp{OBU}. As shown in \cref{fig_model}, adversaries can flood the \ac{VC} systems with fake beacons at high rate (e.g., on the road side with powerful devices) to prevent vehicles from validating other vehicles' beacons in a timely manner. Due to the broadcast nature of \ac{V2V} communication, even a single adversary could create devastating effect as fictitious beacons would affect all vehicles within communication range. Even more so, an adversary could mount \ac{DoS} attacks at multiple points with several devices, resulting in a \ac{DDoS} attack. This can affect the awareness of neighboring vehicles and a number of \ac{VC} applications. The fictitious beacons will be proven invalid when the signatures are verified. However, in a traditional scheme (in the absence of DoS-resilient and efficient verification mechanism), this could significantly delay verifications of authentic beacons, because long waiting times are needed when the queues are saturated with fictitious beacons. More specifically, vehicles cannot distinguish authentic beacons among the received (authentic and fictitious) beacons and verify them first in order to gain awareness of surrounding vehicles.}

\subsection{Security and privacy requirements}

Our scheme is designed to address the above mentioned adversary model taking into consideration the following security and privacy requirements: 

\emph{Authentication and integrity} - Node messages should allow their receivers to corroborate the legitimacy of their senders and verify they were not modified or replayed. We do not require strict identification of the sender, but require the validation that the sender is a legitimate participant of the \ac{VC} systems.

\emph{Non-repudiation and accountability} - Any node can be tied to its actions, and, if need arises, be held accountable and possibly have its long-term identity revealed and have itself evicted from the system.

\emph{Anonymity/Pseudonymity and unlinkability} - A vehicle's beacons should be only linkable over a protocol selectable period, $\tau$. Anonymity/Pseudonymity should be conditional, allowing the system to identify a misbehaving node and evict it.

\emph{\ac{DoS}-resilience} - Nodes should be resilient to adversarial nodes and their flooded fictitious beacons. Even under \ac{DoS} attacks, increase in beacon validation delays should be moderate and nodes should be able to gain awareness of neighboring vehicles in a timely manner.

\begin{table}[h]
	\caption{Notation}
	\centering
	\begin{tabular}{l | *{1}{c} r}
		\hline \hline
		$N$ & \emph{Neighbor size} \\\hline
		$PC$ & \emph{Pseudonymous certificate} \\\hline
		$\{msg\}_{\sigma_{_{PC}}}$ & \emph{Signed message with \ac{PC} attached} \\\hline
		$\alpha$ & \emph{No. of verification results in a beacon} \\\hline
		$\gamma$ & \emph{Beacon frequency} \\\hline
		$\tau$ & \emph{\ac{PC} lifetime} \\\hline
		$H()/H$ & \emph{Hash function/Hash value} \\\hline
		$MAC_{K}(msg)$ & \emph{$H(K || msg)$} \\\hline
		$L_{H}$ & \emph{Hash/MAC size} \\\hline
		$L_{TESLA}$ & \emph{Length of TESLA key chain} \\\hline
		$t_{now}$ & \emph{Fresh timestamp (current time)} \\\hline
		$t_{next}$ & \emph{Next own beacon point after $t_{now}$} \\\hline
		$Pr_{loss}$ & \emph{Probability of packet loss} \\\hline
		\hline
	\end{tabular}
	\renewcommand{\arraystretch}{1}
	\label{table:notation}
\end{table}

\section{{\color{blue}Our scheme}}
\label{sec:scheme}

\subsection{Overview}

Our scheme extends the traditional \ac{V2V} message verification, leveraging cooperating vehicles (referred as \emph{nodes} in the rest of the paper) to defend against \ac{DoS} attacks and reduce validation delays. The basic idea is to augment each (safety) beacon with brief identifiers of previously validated beacons, and attach with TESLA key and MAC. The identifiers indicate the corresponding beacons have been verified by the sender. This is exactly where nodes can benefit from each other: accepting a beacon can help verifying the (received and queued) beacons the identifiers in this beacon point to. The TESLA keys and MACs can expedite message validation, thus remaining resilient to extreme network situations that vehicles receive more beacons than they could handle/verify. Moreover, under a DoS attack, these identifiers point nodes to the potentially valid beacons attached with non-cached \acp{PC}. These beacons are assigned higher priority in the queue so that nodes can obtain faster the sought awareness of surrounding benign nodes. Table~\ref{table:notation} summarizes notation used in this paper.

\begin{figure*}[t]
	\centering
	\includegraphics[width=0.7\textwidth]{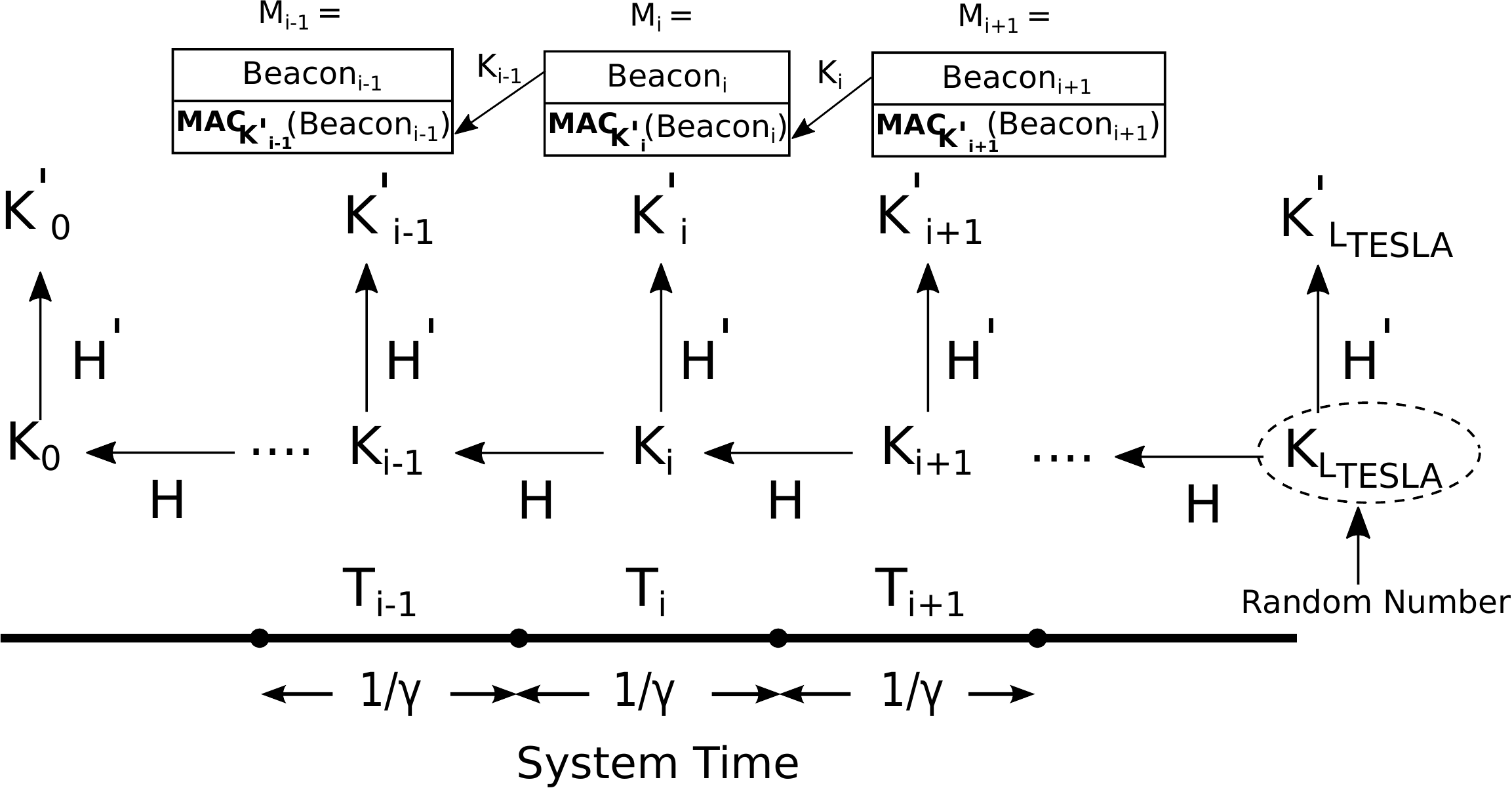}
	\caption{TESLA key chain generation and usage.}
	\label{fig_tesla}
\end{figure*}

\subsection{TESLA key chain generation}

Each node broadcasts safety beacons at a rate $\gamma$. In this paper, we assume $\gamma = 10$ $Hz$, a typical value in the literature~\cite{cam, lyu2016pba, hsiao2011flooding}. Thus, a \ac{PC} can be used for authenticating at most $L_{TESLA} = \tau \cdot \gamma$ beacons. For each pseudonym, a node generates a TESLA key chain~\cite{perrig2000efficient} with a length of $L_{TESLA}$, as shown in \cref{fig_tesla}. $H()$ and $H^\prime()$ are two different hash functions. $K_i$ is the key that will be disclosed with the beacon broadcasted during time slot, $T_{i+1}$, and $K^\prime_i = H^\prime(K_i)$ is the key used to calculate TESLA MAC for the beacon during $T_{i}$. An independent TESLA key chain is generated for each \ac{PC}, in order to ensure unlinkability among the beacons broadcasted (and authenticated) under different \acp{PC}. For example, consider $\tau = 5$ $min$, then a node generates a hash chain for each pseudonym, with $L_{TESLA} = 3,000$. In our scheme, each beacon is signed by the sender with the private key corresponding to the currently valid \ac{PC} and also attached with a TESLA MAC calculated based on the corresponding TESLA key. Each TESLA key is used to authenticate a beacon broadcasted during its corresponding time slot. We assume clocks of nodes are loosely synchronized and system time is split in a same manner for each node. Each time slot ($T_{i}$) is $0.1$ $sec$ and we mandate each node should broadcast only one beacon within a time slot. The exact time points each node broadcasts beacon in a time slot could vary, but we assume beacons from a single node is periodical (i.e., broadcasted every $0.1$ $sec$). The corresponding TESLA key, $K_i$ (and $K^\prime_i$), should be only used to authenticate a beacon broadcasted within the corresponding time slot $T_{i}$. For example, $K^\prime_i$ should be used to generate MAC for $Beacon_i$ and $K^\prime_{i+2}$ should be used to generate MAC for $Beacon_{i+2}$. In case a beacon was skipped or not successfully broadcasted during, e.g., $T_{i+1}$, then $K_{i+1}$ (and $K^\prime_{i+1}$) should be skipped. This helps receivers to authenticate TESLA MACs with correct TESLA keys based on the beacon reception times.

A node needs at most $864,000$ TESLA keys per day (i.e., $24$ $h$) with $\gamma=10$ $Hz$. Thus, the maximum storage needed is $864,000*L_{H}$. For example, with $L_{H} = 80$ $bit$ (such a hash size is sufficient for safety beacons, due to their ephemeral nature), the storage size is $8.64$ $MB$. A node could store a TESLA key every, e.g., 10 slots， and calculate the TESLA keys in between on-the-fly when they are needed. For example, $K_{i}$ and $K_{i+10}$ can be stored, and $K_{i+1}$ ... $K_{i+9}$ can be easily calculated based on $K_{i+10}$; this significantly decreases the storage overhead for TESLA keys. For an off-the-shelf vehicular \ac{OBU}, $864,000$ hash computations can be completed within a few seconds and a storage overhead of $8.64$ $MB$ ($864$ $KB$ if keys at every 10 slots are stored) per day is acceptable.

{\color{blue}

\subsection{Beacon queue maintenance}

The format of a signed beacon in our scheme is changed into:
\begin{flalign}
	\begin{split}
		&Beacon_i = \{Status, t_i, K_{i-1}, H_1, ..., H_\alpha\}_{\sigma_{_{PC}}}\ 
	\end{split}
\end{flalign}

and the format of a beacon message (i.e., a signed beacon attached with a TESLA MAC) is:

\begin{flalign}
	\begin{split}
		&M_i = \{Beacon_i, MAC_{K_i^\prime}(Beacon_i)\}
	\end{split}
\end{flalign}
	
$Status$ consists of vehicle status information, including location, velocity, direction, etc. $t_i$ is the timestamp of the beacon message, which should be within the time slot $T_{i}$. $K_{i-1}$ is the TESLA key for the previous beacon (or time slot). $H_1, ..., H_\alpha$ are the hashes of latest (based on the times of reception) verified beacons: the beacons that the signatures were verified, not cooperatively verified or verified based on TESLA keys and MACs.

\begin{algorithm*}[t]
	\caption{Beacon reception}
	\label{alg:reception}
	\small 
	\begin{algorithmic}[1]
		\State Received a beacon message $M_i$
		\State $M_i = \{Beacon_i = \{Status, t_i, K_{i-1}, H_1 ... H_\alpha\}_{\sigma_{_{PC}}}, MAC_i\}$
		
		\If {$PC \in CACHED\_PC$}
		\If {No beacon with $K_{i-1}$ was received, \textbf{and} $K_{i-1}$ is the correct key corresponds to time slot of $t_i$}
		\State Insert $\{M_i, H(M_i)\}$ to the head of $Queue_1$.
		\State Find beacon message, $M^\prime$, attached with $PC$ from $Queue_1$.
		\If {$M^\prime$ is found}
		\State Remove $M^\prime$ from $Queue_1$.
		\State Input $M^\prime$ to \cref{alg:tesla} for TESLA MAC validation.
		\EndIf
		\Else
		\State {Drop $M_i$.}
		\EndIf
		\Else
		\State Insert $\{M_i, H(M_i)\}$ to the head of $Queue_1$.
		\EndIf
	\end{algorithmic}
\end{algorithm*}

\begin{algorithm*}[h!]
	\caption{Queue element selection}
	\label{alg:queue}
	\small 
	\begin{algorithmic}[1]
		\If {$Queue_2$ is not empty}
		\State Choose first element, $E$, from $Queue_2$. 
		\If {Multiple elements exist in $Queue_2$ with same $PC$ of $E$}
		\State Choose the element, $E$, with the latest timestamp.
		\EndIf
		\ElsIf {$Queue_1$ is not empty}
		\State $k$ element(s) exist(s) in $Queue_1$ with beacon timestamps satisfy $t_{beacon} + 1/\gamma > t_{next}$ .
		\If {$k > 0$}
		\State Uniform randomly selects an element, $E$, from $k$ first elements of $Queue_1$.
		\ElsIf {$k = 0$}
		\State Choose first element, $E$, from $Queue_1$.
		\EndIf 
		\EndIf
		\State Input $E$ to \cref{alg:coop}.
	\end{algorithmic}
\end{algorithm*}

\cref{alg:reception} shows the beacon reception process. Consider a received beacon message, $M_i$. Attached \ac{PC} of $M_i$ is checked first. If it was \emph{verified and cached} (simplified as \emph{cached} in the rest of the paper), then the receiver checks whether the attached TESLA key is correct. This is done by validating it against the TESLA key in a previous beacon from the same sender\footnote{\color{blue} We refer a pseudonymous identity as a sender. Thus, messages attached with the \ac{PC} are considered sent from the same sender, while beacons sent under two different \acp{PC} (from the same node) can be considered sent from two different senders in this context.}. For example, if a beacon, $M_{i-2}$, received during the time slot, $T_{i-2}$, was verified based on signature and includes a TESLA key, $K_{i-3}$. Then, the correct TESLA key disclosed in $T_{i}$ should be $H^2(K_{i-3})$, i.e., $H(H(K_{i-3}))$. If multiple beacons with the correct TESLA key, $K_{i-1}$, were received, only the first of them is kept, due to the nature of single hop communication (see \cref{sec:analysis} for further analysis).

If $M_{i}$ satisfies the above requirement or the attached \ac{PC} is non-cached, then the receiver generates a queue element and insert it to the head of $Queue_1$:

\begin{align}
\{M_{i}, H(M_i)\}
\end{align}

Each node maintains two queues. $Queue_1$ stores newly received beacon messages and $Queue_2$ stores potentially valid beacon messages with non-cached \acp{PC} attached to them. Elements in $Queue_2$ are extracted from $Queue_1$ during the execution of \cref{alg:coop,alg:tesla} (explained below). $Queue_2$ is given higher priority in order to ensure timely awareness of newly encountered nodes when computation resource is scarce; the beacons signed under the cached \acp{PC} can be validated through the ``cheaper'' MACs.

The selection of the beacons to verify from the queues is done according to \cref{alg:queue}. When $Queue_2$ is not empty, the node pops an element from the head. Consider the sender of this element (i.e., beacon) is $V$. If multiple elements sent from $V$ exist in $Queue_2$, the element with the most recent timestamp is chosen: validating this beacon can further validate earlier beacons based on MACs.

\begin{algorithm*}[t]
	\caption{Cooperative verification}
	\label{alg:coop}
	\small 
	\begin{algorithmic}[1]
		
		\State $E = \{M, H(M)\}$
		\State $M = \{Beacon = \{Status, t_i, K_{i-1}, H_1 ... H_\alpha\}_{\sigma}, MAC_i\}$
		\If {The signature of $Beacon$ is valid}
		\State Accept $Beacon$.
		\If {$PC$ of $Beacon$ is new}
		\State Add $PC$ to $CACHED\_PC$.
		\State Extract beacon(s) signed under the same $PC$ from $Queue_2$ and $Queue_1$ except the latest one;
		\State store the extracted beacon(s) to $BEACON\_SET$ and remove $BEACON\_SET$ from $Queue_1$.
		\For {Each $M_i$ in $BEACON\_SET$}
		\State Input $M_i$ to \cref{alg:tesla} for TESLA MAC validation.
		\EndFor
		\EndIf
		\For {Each $H_i$ in $H_1..H_\alpha$ of $Beacon$}
		\If {An element, $E^\prime$, with the message hash, $H_i$, is found in $Queue_1$}
		\State $E^\prime = \{M^\prime = \{Beacon^\prime, MAC^\prime\}, H(M^\prime)\}$
		\If {$PC^\prime$ of $Beacon^\prime$ is in $CACHED\_PC$}
		\State Accept $Beacon^\prime$.
		\Else
		\State Remove $E^\prime$ from $Queue_1$ and insert to the end of $Queue_2$.
		\EndIf
		\EndIf
		\EndFor
		\EndIf
	\end{algorithmic}
\end{algorithm*}

\begin{figure*}[h!]
	\centering
	\includegraphics[width=\textwidth]{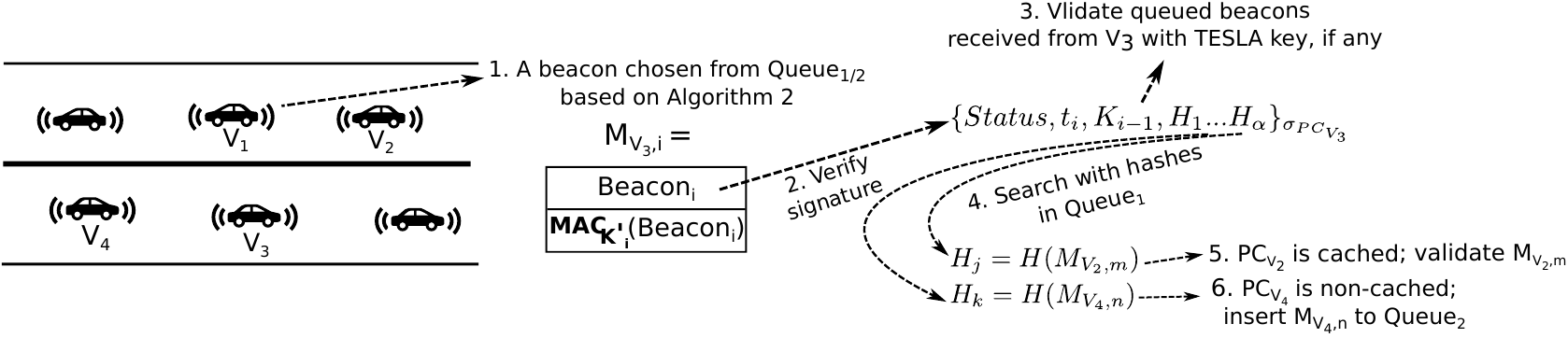}
	\caption{{\color{blue} An example of cooperative verification steps (Algorithm~\ref{alg:coop}).}}
	\label{fig_coop}
\end{figure*}

If $Queue_2$ is empty, an element from $Queue_1$ is selected. Unlike element selection from $Queue_2$, the element here is randomly chosen among those that satisfy the condition: $t_{beacon} + 1/\gamma > t_{next}$, where $t_{beacon}$ is the timestamp of each queued beacon and $t_{next}$ is the time point for the next own beacon broadcast. Recall that in \cref{alg:reception}, the new elements are inserted at the head of $Queue_1$ and, here, an element is chosen again from the head of $Queue_1$. The following three reasons led to such a \ac{LCFS} and conditional randomized selection process:

\begin{itemize}

	\item Verifications of fresh beacons benefit other nodes when the results are shared through own beacons. For example, in a traditional \ac{FCFS} scheme, consider a node ($V_1$) that chooses a beacon, $M_{V_2,i-1}$, from $Queue_1$ for signature verification, while the latest beacon from $V_2$, $M_{V_2,i}$, was lost. Then, sharing this result is not helpful for the receiver, $V_3$, which received both $M_{V_2,i-1}$ and $M_{V_2,i}$, because $M_{V_2,i-1}$ was already validated based on the TESLA key disclosed with $M_{V_2,i}$.
	
	\item Randomized selection among fresh beacons guarantees that the nodes would not choose the same latest received beacon to verify, thus decreases duplicated signature verifications performed by the cooperating nodes.
	
	\item The time condition ($t_{beacon} + 1/\gamma > t_{next}$) guarantees that, before the dissemination of next own beacon, the beacons, corresponding to piggybacked verification results, have not been validated based on TESLA MACs by receivers, otherwise, the shared verification results could be less useful.
	
\end{itemize}

If no beacon message satisfies the time condition, then the first queue element is popped from the head of $Queue_1$.

\begin{algorithm*}[t]
	\caption{TESLA MAC Validation}
	\label{alg:tesla}
	\small 
	\begin{algorithmic}[1]
		\State $M_i = \{Beacon_i = \{Status, t_i, K_{i-1}, H_1 ... H_\alpha\}_{\sigma}, MAC_i\}$
		\If {$MAC_{K^\prime_{i}}(Beacon_i) = MAC_i$}
		\State Accept $Beacon_i$.
		\For {Each $H_i$ in $H_1..H_\alpha$ of $Beacon$}
		\If {An element, $E$, with the message hash, $H_i$, is found in $Queue_1$}
		\If {PC of $Beacon$ is not in $CACHED\_PC$}
		\State Remove $E$ from $Queue_1$ and insert to the tail of $Queue_2$. 
		\EndIf
		\EndIf
		\EndFor
		\Else
		\State Drop $M$.
		\EndIf
	\end{algorithmic}
\end{algorithm*}

\subsection{Cooperative beacon verification}

\cref{alg:coop} shows our cooperative verification scheme, taking $E$ chosen from \cref{alg:queue} as the input. The signature(s) of (the attached non-cached \ac{PC} and) $M$ is(are) verified first. If \ac{PC} is non-cached, then it is cached and the beacons from the same sender can be validated based on TESLA MACs (except the most recent one, for which the corresponding TESLA key has not been disclosed/received). The hashes $H_1,...,H_\alpha$ in $M$ are used to validate further the beacons in $Queue_1$. The node searches, with each $H_i$, in $Queue_1$ for a matching $H(M^\prime)$. If a matching queue element is found, the node checks whether the attached $PC^\prime$ is cached. If yes, then $M^\prime$ is accepted. Otherwise, the element is inserted to the tail of $Queue_2$. For the latter case, $M^\prime$ is not simply accepted because this validation is correlated with the validations of extra beacon messages carrying the same $PC^\prime$ based on TESLA MACs, thus the security risk is high (see Sec.~\ref{sec:analysis} for further analysis). \cref{fig_coop} shows an example of the cooperative verification scheme: $V_1$ chooses a beacon from its queue for signature verification. After the signature verification is passed, $K_{i-1}$ is used to validate queued beacons sent from $V_3$. A piggybacked hash value, $H_j$ is found to match a beacon attached with $PC_{V_2}$. As a result, $M_{V_2,m}$ is accepted, because $PC_{V_2}$ is already cached. Moreover, another hash value, $H_k$, is found to match a beacon, $M_{V_4,n}$, while $PC_{V_4}$ is non-cached. Therefore, $M_{V_4,n}$ is inserted to $Queue_2$ for a signature verification later.

\subsection{TESLA-based validation}

Each TESLA-based validation (in \cref{alg:reception,alg:coop}) is processed according to \cref{alg:tesla}. If the TESLA MAC is validated, then each $H_{i}$ in the validated beacon is used to search matching beacon message from $Queue_1$. If there is any matching beacon attached with non-cached \ac{PC}, then it is inserted to the tail of $Queue_2$. The hashes are only used to search for potentially valid beacon messages attached with non-cached \acp{PC}, but not used for validating beacons, because TESLA-based authentication does not provide non-repudiation. Therefore, the TESLA-validated beacons should not be used to further validate more beacons (see Sec.~\ref{sec:analysis} for further analysis).

}

\section{Privacy and security Analysis}
\label{sec:analysis}

In this section, we provide privacy and security analysis for our scheme. We emphasize that we are not concerned here with the validity of the message content, e.g., the correctness of a location or an alert about emergency braking; those are orthogonal and can be addressed by relevant consistency checking~\cite{festag2010design, leinmuller2006improved} and data-centric security~\cite{raya2008data} schemes. Here, we are concerned with incorrectly signed (with arbitrary content) messages, and the attempt to saturating benign vehicles with (fake) signature verifications while affecting validations of authentic safety beacons.

{\color{blue}

\subsection{Privacy analysis}

We note that privacy is not weakened by our scheme. The shared verification results in a beacon do not link transmissions of any other node, beyond what one can infer from the geographical information included in the beacons themselves. An independent TESLA key chain is generated for each \ac{PC}, thus messages authenticated under different \acp{PC} cannot be linked. Beacons correlated (i.e., linkable) based on the TESLA keys from a same TESLA key chain are also signed under the same \ac{PC}, so that they can be already trivially linked based on the \ac{PC} (at most for a period, $\tau$) even without TESLA keys, as is the case for a traditional scheme.

\subsection{Security analysis}

Beacons in our scheme are validated based on one of three components: (1) signature, (2) MAC, and (3) shared verification result. We first analyze security properties for each validation approach. Then, we discuss non-repudiation provision, notably how to compensate for the lack of non-repudiation in TESLA. Finally, we conclude with the analysis of resilience to DoS attacks of our scheme.

\textbf{Signature verification:} The use of pseudonymous authentication, as per the standards under development~\cite{cam,khodaei2018secmace}, guarantees non-repudiation and message integrity and authentication; as long as the receiving node performs the cryptographic validation itself (message signature and attached pseudonym validation). Inclusion of fresh timestamps in messages and timestamp checking prevent replay of messages.

\textbf{TESLA-based validation:} TESLA provides message integrity and authentication after hash chain anchors are verified with signatures~\cite{perrig2000efficient}. Time synchronization (e.g., through GPS) guarantees that only the corresponding TESLA keys are used at any point in time (see \cref{fig_tesla}). Our scheme prevents memory exhaustion attacks, because at most one beacon carrying the correct TESLA key needs to be queued. This is due to the nature of single-hop communication. The first beacon with the correct TESLA key would be either (1) the (potentially) authentic one, or, (2) if the authentic one was lost, the first among the masqueraded beacons attached with the correct TESLA key (overheard by adversary after it was disclosed by its legitimate sender). However, the latter one will be proven invalid either through its signature verification or TESLA MAC validation. This is not more harmful than receiving a randomly created fake beacon, which is even easier to generate from the perspective of an adversary. We mandate that each beacon is signed by the sender. If the content of a beacon is suspicious, the receiver can choose to verify the signature on the beacon, thus non-repudiation can be achieved.

\textbf{Cooperative verification:} Shared verification results in beacon messages are used to either find potentially valid beacon messages attached with non-cached \acp{PC} or validate latest beacons attached with cached \acp{PC}. For a cached \ac{PC}, there is always at most one beacon message in $Queue_1$ from the same sender. A shared verification result matching this beacon could be used to validate this beacon. In this case, non-repudiation, authentication and integrity are not directly achieved based on signature or TESLA MAC on the validated beacon, but the trust on this beacon is established leveraging another node. If the validated beacon is proven fake later, the node that shared this verification result is accountable for this misbehavior. A newer beacon from the same sender (as the validated beacon) can help further corroborate the correctness of cooperatively validated beacon (based on TESLA MAC). Moreover, with mobility prediction~\cite{ristanovic2011adaptive} and content validation~\cite{festag2010design, leinmuller2006improved} approaches, a vehicle can choose to accept the cooperatively validated beacon only when the beacon content is within the prediction/validation error threshold. Again, when the content of any beacon is suspicious, signature can be verified to achieve immediate non-repudiation, authentication and integrity.

On the other hand, for a non-cached \ac{PC}, there could exist several beacon messages from the same sender in $Queue_1$. A falsely accepted fake \ac{PC} could result in accepting a series of false beacons. Therefore, the signature of a beacon message attached with non-cached \ac{PC} should be verified (by inserting the beacon message to $Queue_2$) before the correlated beacons can be validated based on TESLA MACs.

\textbf{Non-repudiation:} We mandate each beacon be signed by the sender, so that signature verification can ensure non-repudiation. However, TESLA-based validation and cooperative verification cannot guarantee that the validated/accepted beacons are properly signed. Although accountability for cooperatively verified beacons can be traced to senders of verification results; TESLA-based validation does not provide non-repudiation, as validations are merely based on symmetric TESLA keys.

Consider an adversary fabricates beacons with disclosed TESLA keys and claim the beacons were received at correct time slots (with correct TESLA keys and fake signatures). The original generator of those TESLA keys cannot deny the fabricated beacons were not sent by him-/her-self. On the other hand, with this in mind, a legitimate sender can also send beacons with correct TESLA keys and fake signatures at correct time slots. A receiver would accept the beacons with TESLA-based validations, and realize the signatures were invalid later. However, as TESLA-based validations do not provide non-repudiation, the sender cannot be held accountable, exactly because the beacons presented by the victim might have been fabricated (by the victim the same way the adversary did).

With such a vulnerability, TESLA-based validations should be carefully used. This is the case in our scheme: TESLA-based validations are only used for validating successive beacons attached with cached \acp{PC}. Redundancy of frequent beacons and predictability of vehicle status within a short period~\cite{nguyen2017mobility} can minimize the vulnerability incurred by the lack of non-repudiation: a beacon that deviates too much from a previous signature-validated beacon can be considered suspicious and the signature of this beacon can be verified as a backup approach. Moreover, the piggybacked hash values in TESLA-validated beacons are only used for discovering potentially valid beacons from new senders, which are verified based on signatures later.

}

\textbf{Thwarting clogging (D)DoS:} With TESLA-based validation, once a TESLA key for a \ac{PC} is stored after a successful signature verification, then the rest of the beacons form the same sender can be validated based on TESLA MACs. However, TESLA-based validation could be helpful only after a successful signature verification for a beacon message attached with a non-cached \ac{PC}. Discovery of authentic beacons attached with non-cached \acp{PC} still remains an issue. This is problematic especially when vehicles reach a time point for pseudonym change. Flooded fictitious beacons can affect the verifications of new \acp{PC}, thus the awareness of surrounding vehicles.

With our cooperative verification scheme, vehicles can efficiently fetch potentially valid beacons attached with non-cached \acp{PC} based on shared verification results and verify them first in order to gain awareness of newly encountered vehicles with short delays. Once a beacon from a sender vehicle has been successfully verified, the successive beacon messages from the same sender can be validated based on TESLA MACs or shared verification results.

\section{Performance evaluation}
\label{sec:simulation}

In this section, we evaluate our scheme with simulations. We show that our scheme achieves low delays in high vehicle density scenario, and even under DDoS attacks, which would not have been possible for the standard approach (signature verification of all received beacon messages, referred as \emph{baseline} scheme in the rest of the paper).

\begin{table}[t]
	\caption{System Parameters (\textbf{\emph{Bold}} for Default Setting)}
	\centering
	\renewcommand{\arraystretch}{1.10}
	\begin{tabular}{ l | c }
		\hline \hline
		$N$ & 20, 30, 40, 50, \textbf{\emph{60}}, 70, 80 \\\hline
		$\gamma$ & \textbf{\emph{10}} $Hz$ \\\hline
		$T_{vrfc}$ & \textbf{\emph{4}} $ms$ \\\hline
		$Pr_{loss}$ & 0.1, \emph{\textbf{0.2}}, 0.3, 0.4, 0.5, 0.6, 0.8 \\\hline
		$\alpha$ & 1, 2, 3, \textbf{\emph{4}}, 5  \\\hline
		$N_{adv}$ & \textbf{\emph{4}} (Static), \textbf{\emph{10}} (Mobile) \\\hline
		$\gamma_{adv}$ & \textbf{\emph{250}} $Hz$ (No packet loss) \\\hline	
		\hline
	\end{tabular}
	\renewcommand{\arraystretch}{1}
	\label{table:parameter}
\end{table}

\subsection{Simulation settings}

We use OMNeT++~\cite{omnet} to simulate our scheme and analyze the system performance. We consider two mobility scenarios: a static scenario and a mobile scenario. For the mobile scenario, Veins~\cite{sommer2011bidirectionally} is used for the connection between SUMO~\cite{SUMO2012} and OMNeT++. Table~\ref{table:parameter} shows the system parameters and values used in the simulation. We assume a communication range of $200$ $m$ with a packet loss ratio of $Pr_{loss}$, i.e., a node can successfully receive a broadcasted beacon within the communication range with a probability, $1-Pr_{loss}$. We assume a signature verification delay of $T_{vrfc}$ for both \ac{PC} verification and message verification. For simplicity, we assume the beacon validations based on hash computations (i.e., cooperative validations or TESLA-based validations) incur zero delay (in reality, they introduce a tiny delay, which can be in the order of $\mu s$). For each simulation setting, we perform 5 randomly seeded experiments of 1 $min$ and the results are averaged over these 5 runs. The bold values in Table~\ref{table:parameter} are the default ones used in our simulation. For example, when $N$ is the parameter we examine (\cref{subfig_static_waiting_node}), then the rest of the parameters have the default values: $Pr_{loss}=0.2$, $T_{vrfc}=4$ $ms$, $\alpha={4}$ and $\gamma=10$ $Hz$. $N_{adv} = 4$ and $\gamma_{adv} = 250$ $Hz$ are only set for adversarial network scenarios. The default values of $T_{vrfc}$ and $\gamma$ are typical values based on the literature (e.g.,~\cite{calandriello2011performance,cam}). Note that we assume all beacons from adversaries are received by benign nodes (within the communication range without any packet loss) in order to simulate an extreme situation. For example, when $Pr_{loss} = 0.8$, the equivalent real beacon rate from each adversarial node would be $1250$ $Hz$. This could be non-realistic for a single adversary, but we can consider one adversarial node in our evaluation as an aggregate of multiple adversarial nodes in reality.

\subsection{Static mobility scenario}

\begin{figure}[t]
	\centering
	\includegraphics[width=0.6\columnwidth]{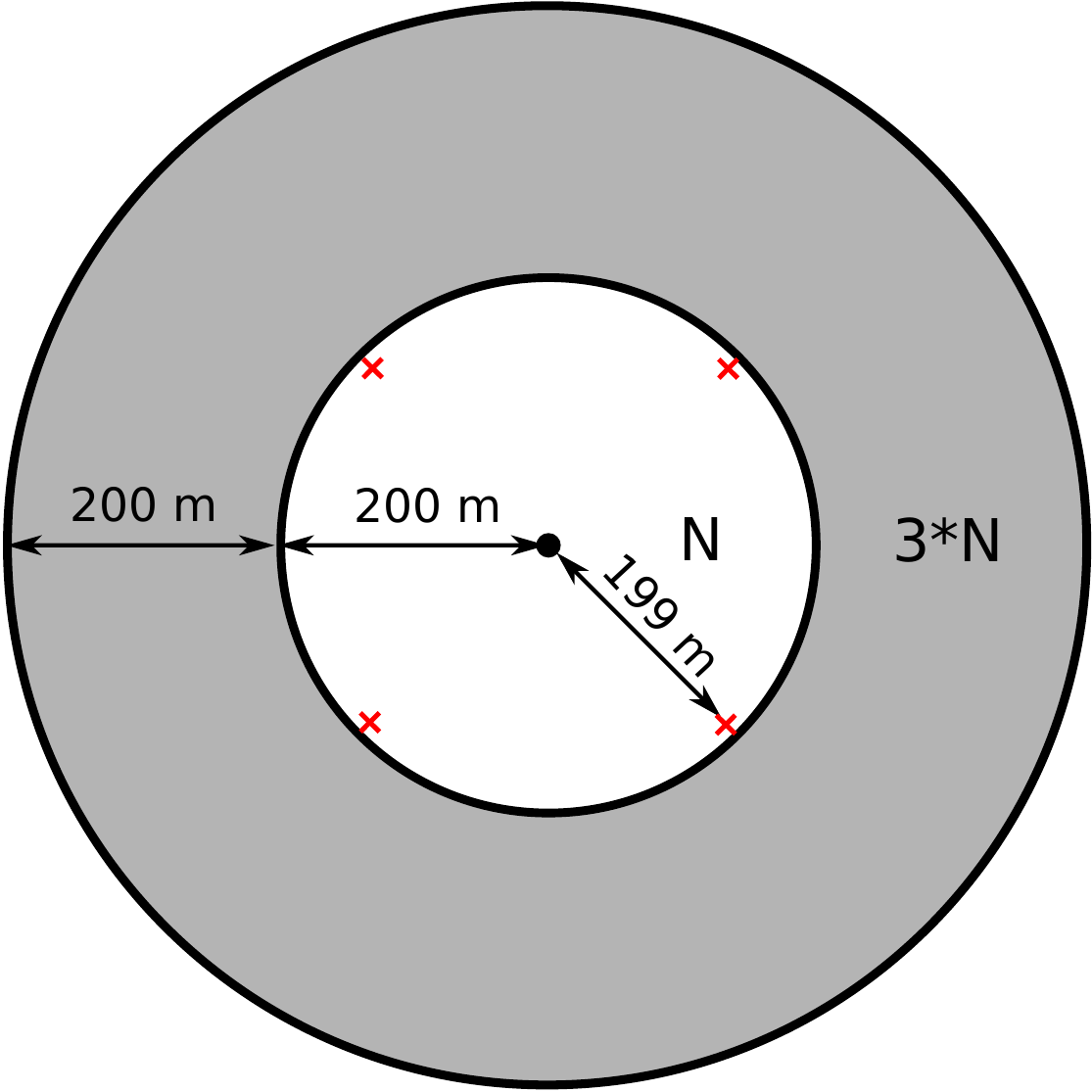}
	\caption{Static scenario: Simulated area with the evaluated node at the center with a neighbor density of $N$. X marks indicate placement of adversarial nodes.}
	\label{fig_static}
\end{figure}

\cref{fig_static} shows the simulated area for the static scenario: we place the node we evaluate in the center of a $200$ $m$ disc area, and place $N$ nodes randomly within the disc. Moreover, another $3*N$ nodes (thus, a same node density as in the inner disc) are placed at the outer disc (i.e., gray area). {\color{blue} We consider this setup to avoid evaluating the scheme in a setting that is overly optimistic and favorable for our scheme.	As we assume an effective communication range of 200 $m$, the node we evaluate receives beacons from senders within the inner disc. However, the node in the inner disc will also receive messages from the outer disc (gray area). We add this $200\ m$ extra area to emulate a more realistic scenario, in which the evaluated node could also receive verification results that it is not interested in. Otherwise, if simulated only with the inner disc, the scenario would be very optimistic: all the verification results piggybacked by the received beacons will be useful (i.e., corresponds to beacons within the communication range).	
}

\begin{figure*}[t]
	\centering
	\begin{subfigure}{.24\textwidth}
		\includegraphics[width=\columnwidth]{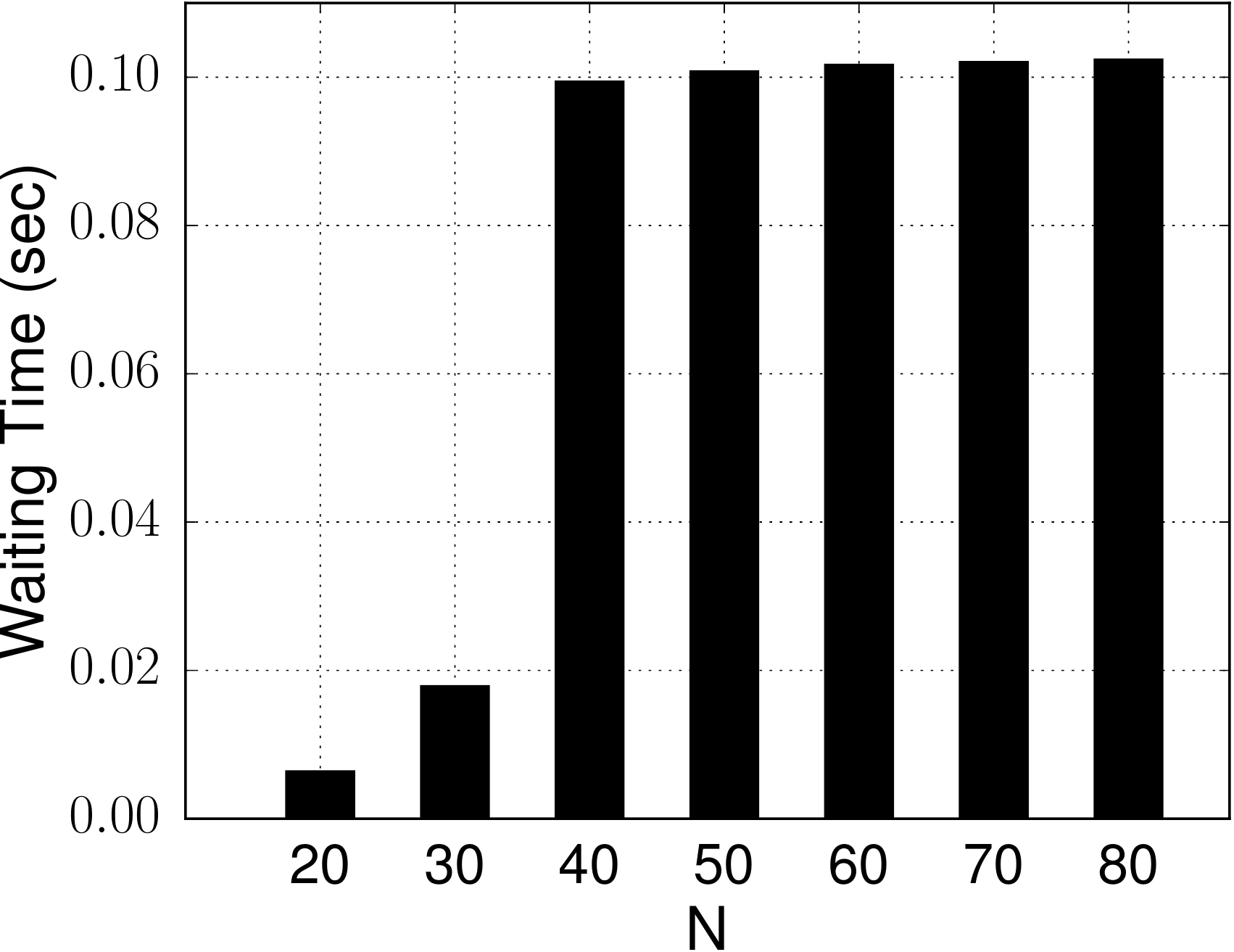}%
		\caption{}%
		\label{subfig_static_waiting_baseline}%
	\end{subfigure}\hspace{1mm}
	\begin{subfigure}{.24\textwidth}
		\includegraphics[width=\columnwidth]{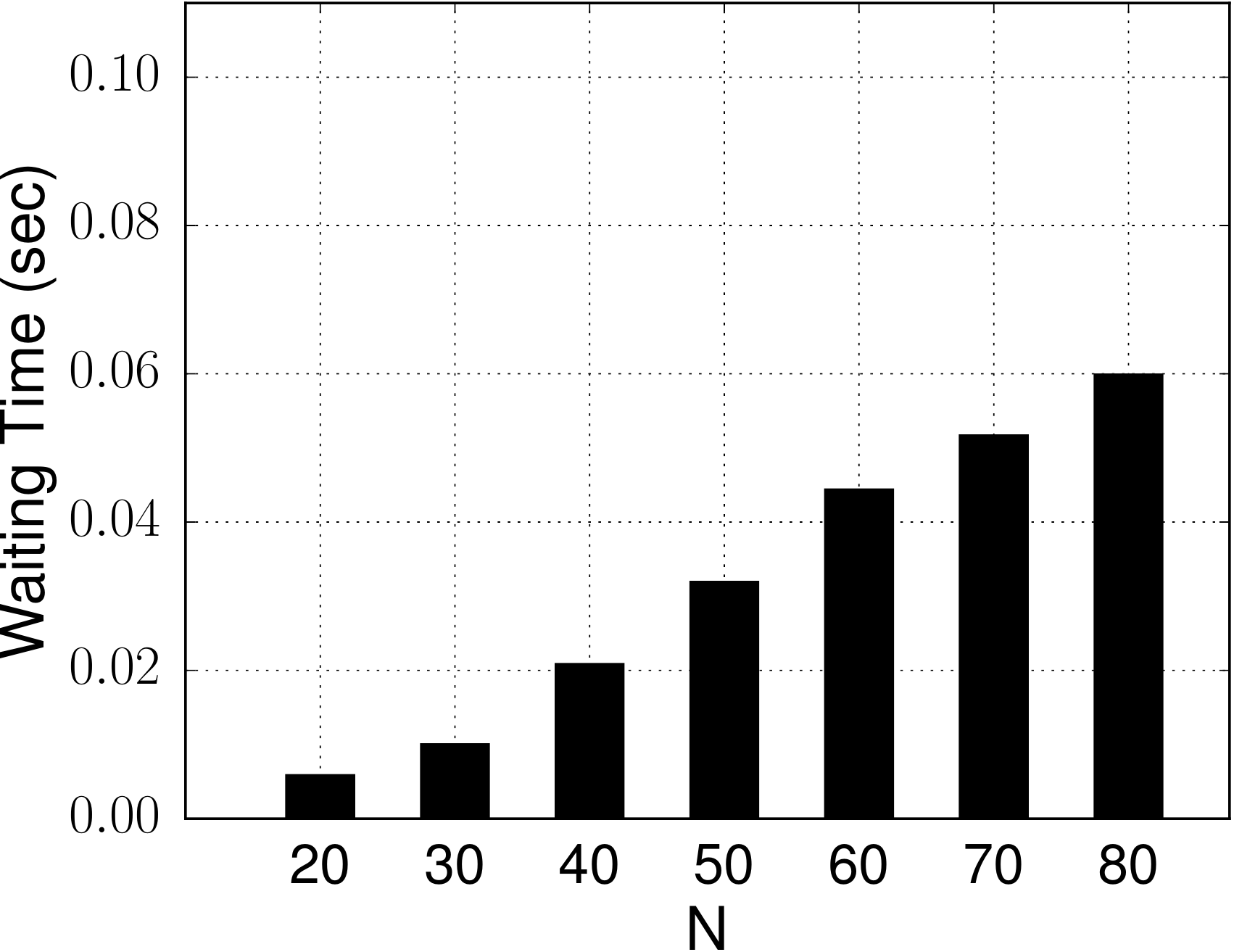}%
		\caption{}%
		\label{subfig_static_waiting_node}%
	\end{subfigure}\hspace{1mm}
	\begin{subfigure}{.24\textwidth}
		\includegraphics[width=\columnwidth]{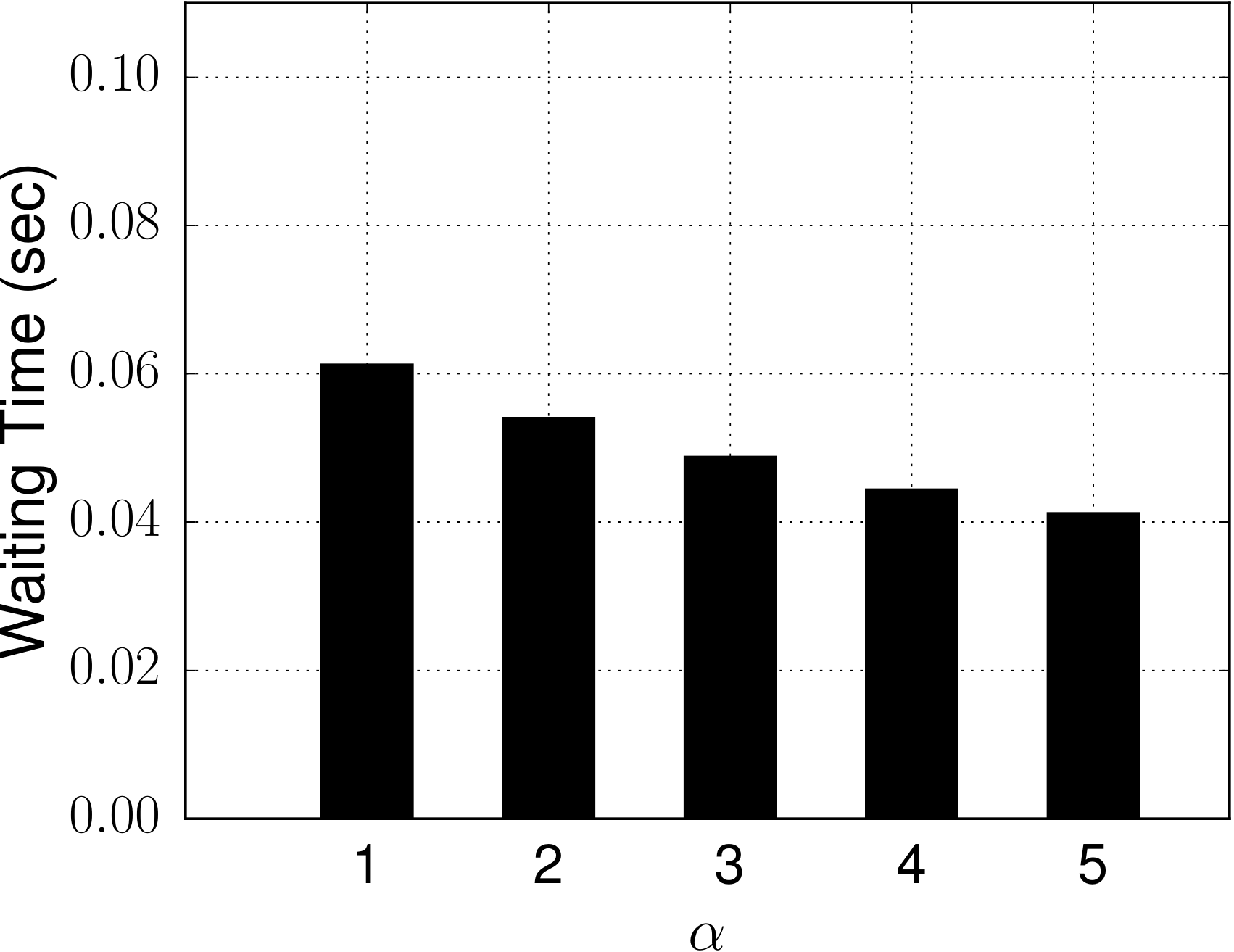}%
		\caption{}%
		\label{subfig_static_waiting_num}%
	\end{subfigure}\hspace{1mm}
	\begin{subfigure}{.24\textwidth}
		\includegraphics[width=\columnwidth]{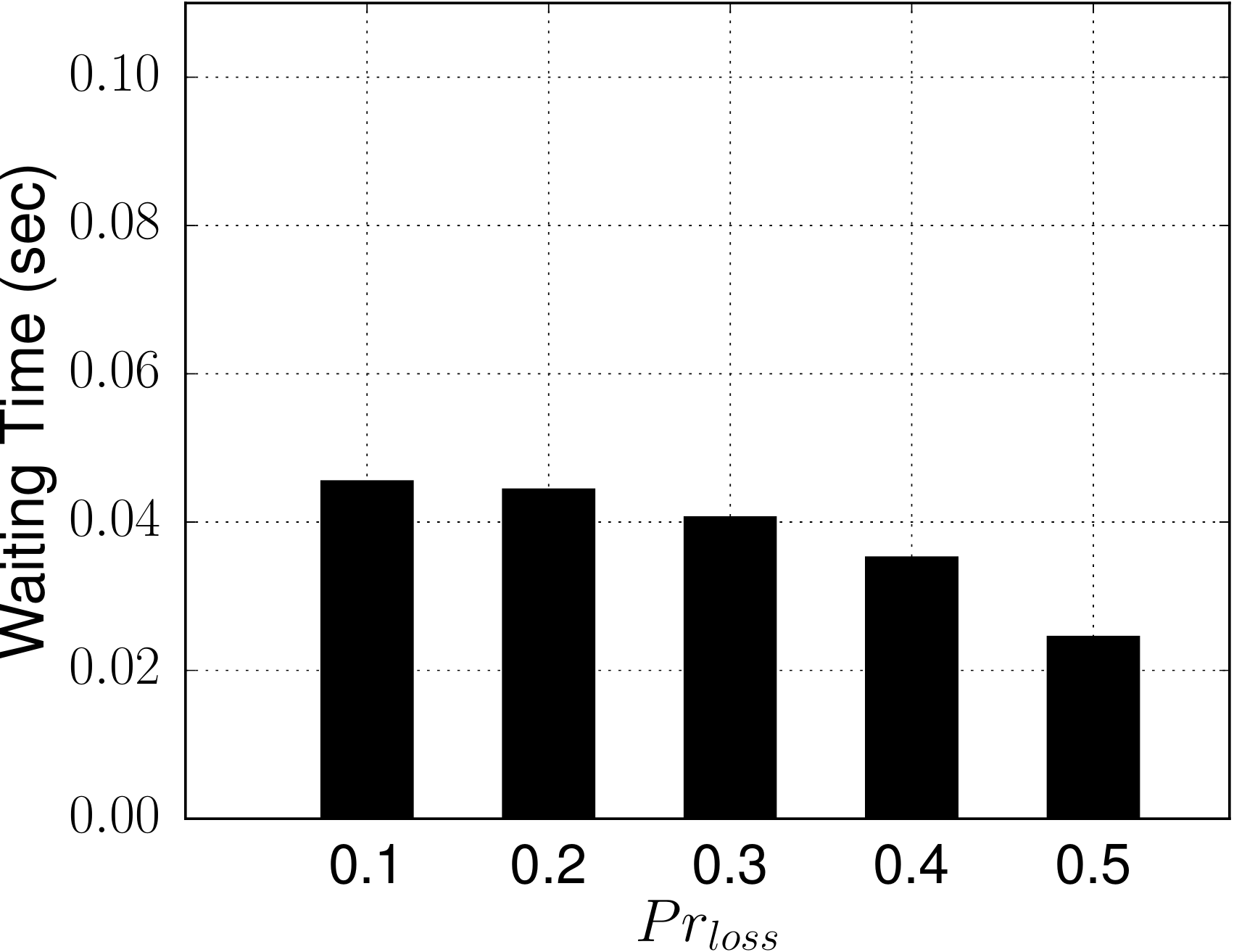}
		\caption{}%
		\label{subfig_static_waiting_loss}%
	\end{subfigure}
	\caption{Benign and Static: Average waiting time as a function of $N$ with (\subref{subfig_static_waiting_baseline}) baseline scheme ($\alpha = 0$ and TESLA for $N \geq 40$) and (\subref{subfig_static_waiting_node}) cooperative beacon verification, and as a function of (\subref{subfig_static_waiting_num}) $\alpha$ and (\subref{subfig_static_waiting_loss}) $Pr_{loss}$ under default settings.}
	\label{fig_static_waiting}
\end{figure*}
\begin{figure*}[h]
	\centering
	\begin{subfigure}{.24\textwidth}
		\includegraphics[width=\columnwidth]{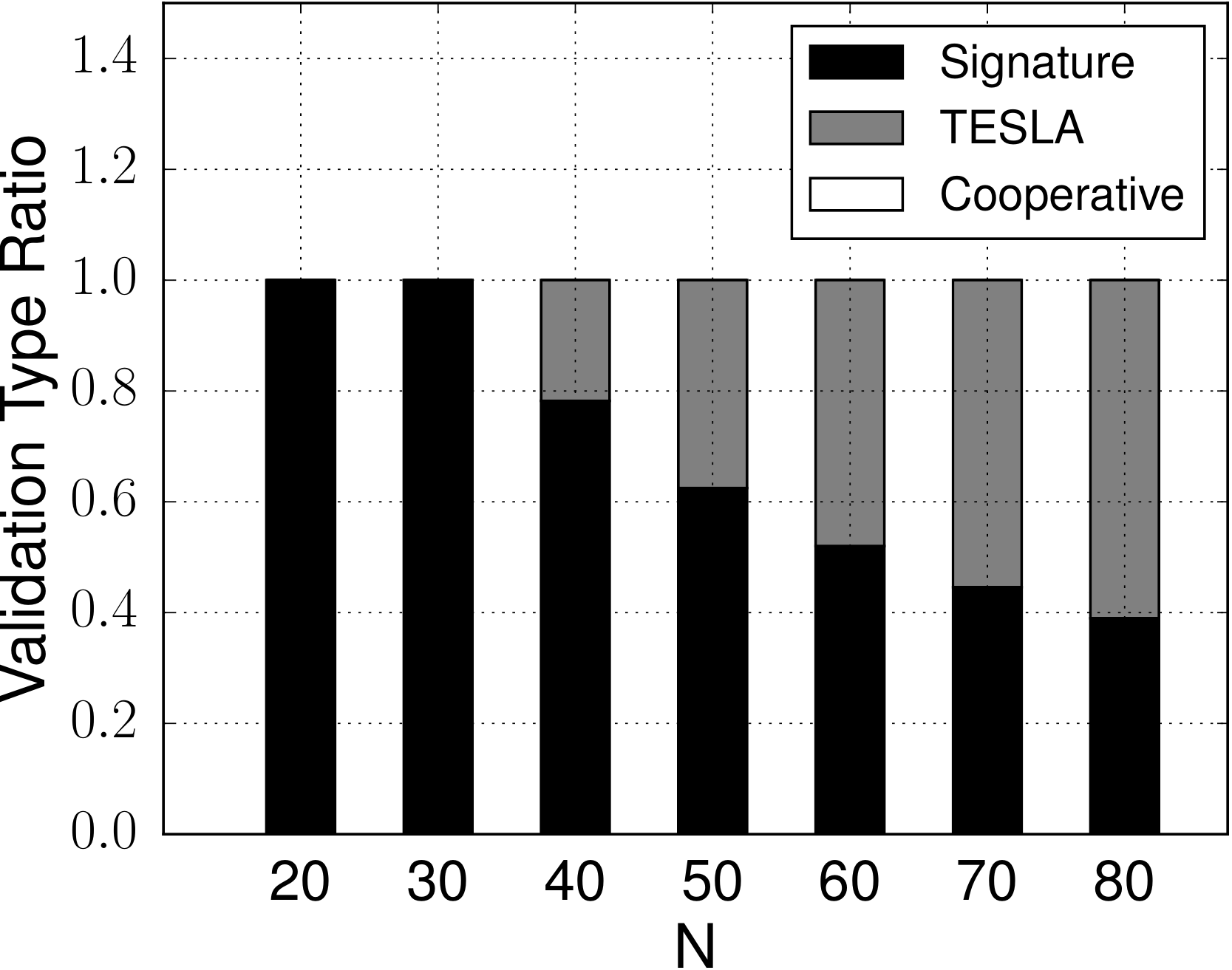}%
		\caption{}%
		\label{subfig_static_type_baseline}%
	\end{subfigure}\hspace{1mm}
	\begin{subfigure}{.24\textwidth}
		\includegraphics[width=\columnwidth]{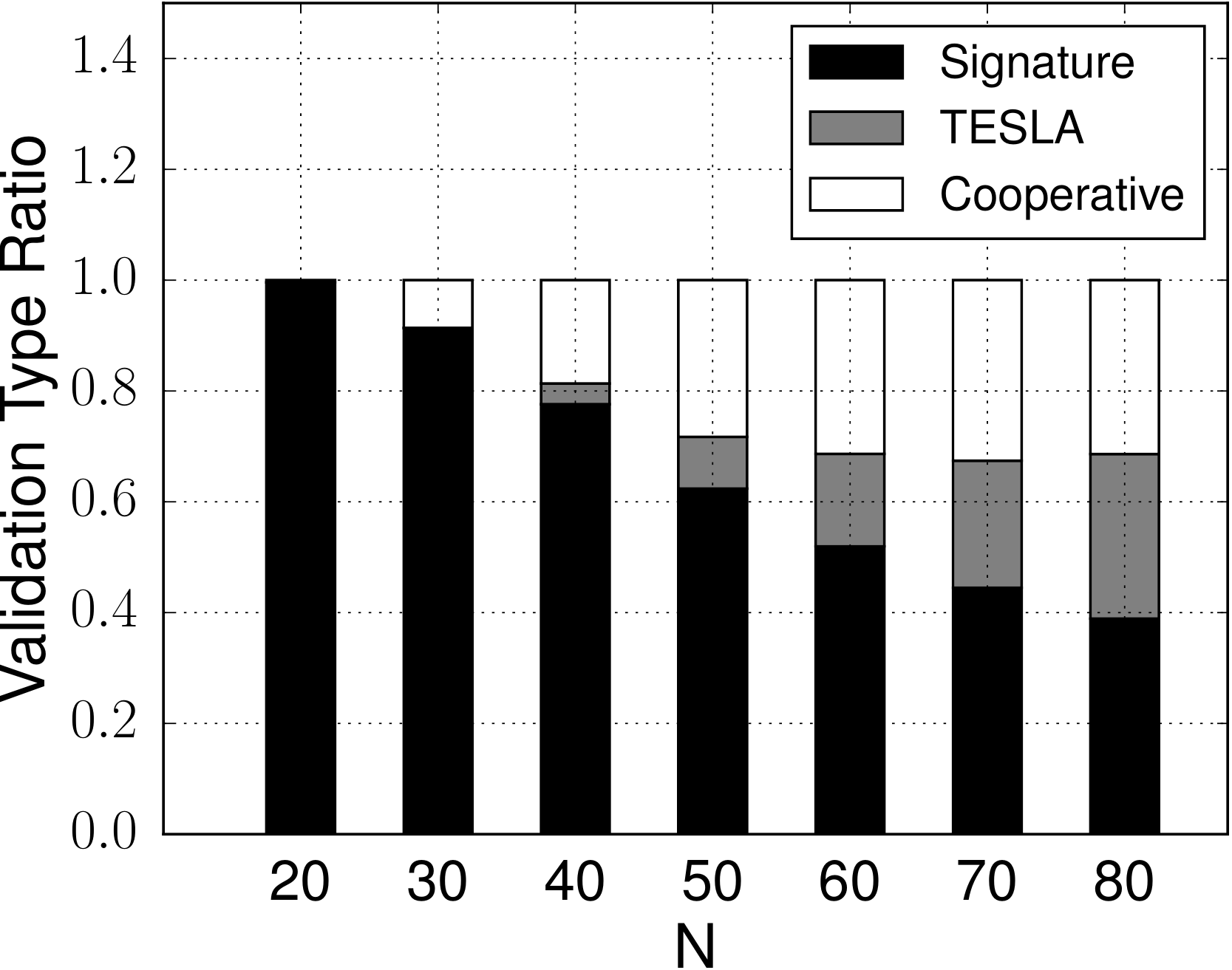}%
		\caption{}%
		\label{subfig_static_type_node}%
	\end{subfigure}\hspace{1mm}
	\begin{subfigure}{.24\textwidth}
		\includegraphics[width=\columnwidth]{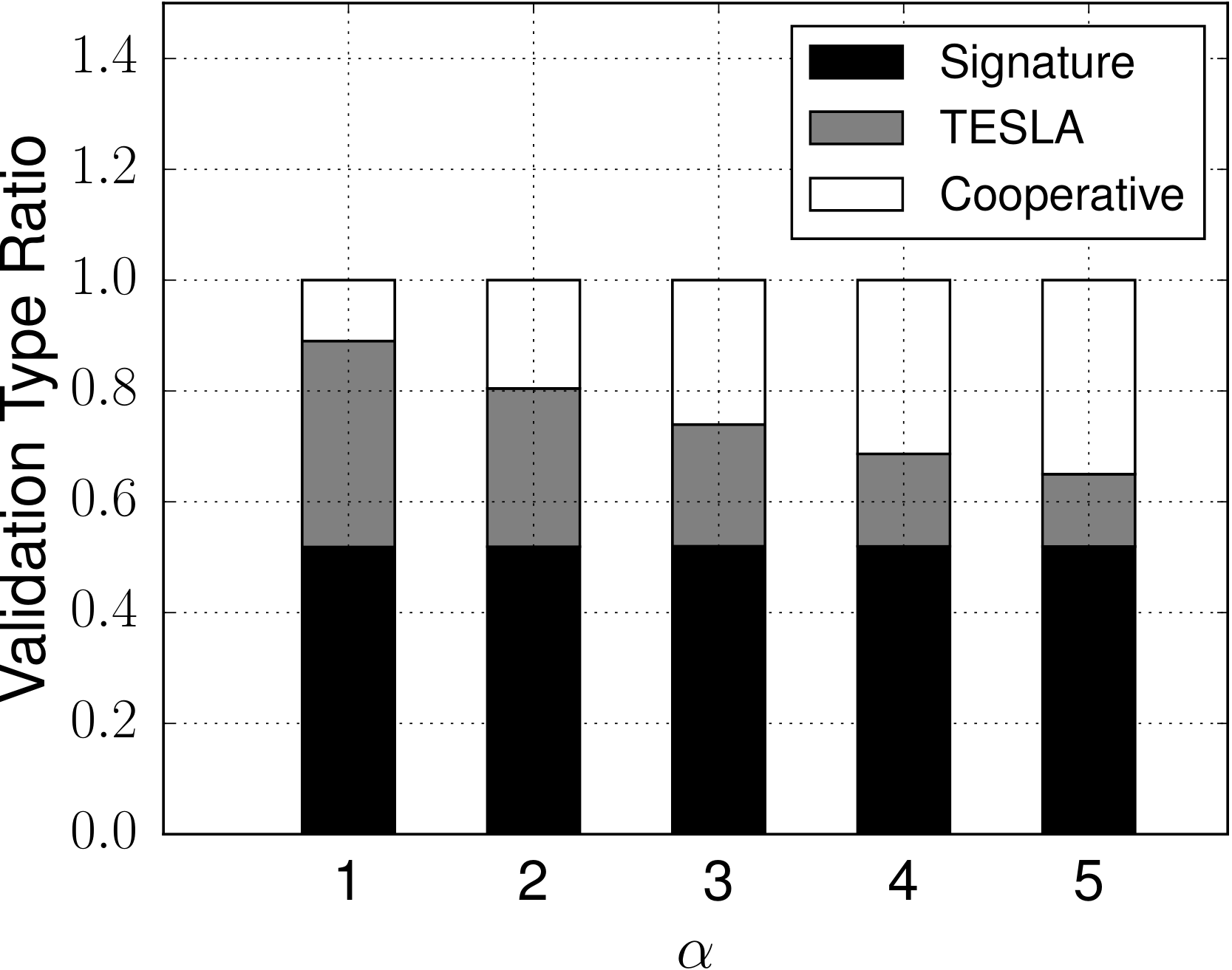}%
		\caption{}%
		\label{subfig_static_type_num}%
	\end{subfigure}\hspace{1mm}
	\begin{subfigure}{.24\textwidth}
		\includegraphics[width=\columnwidth]{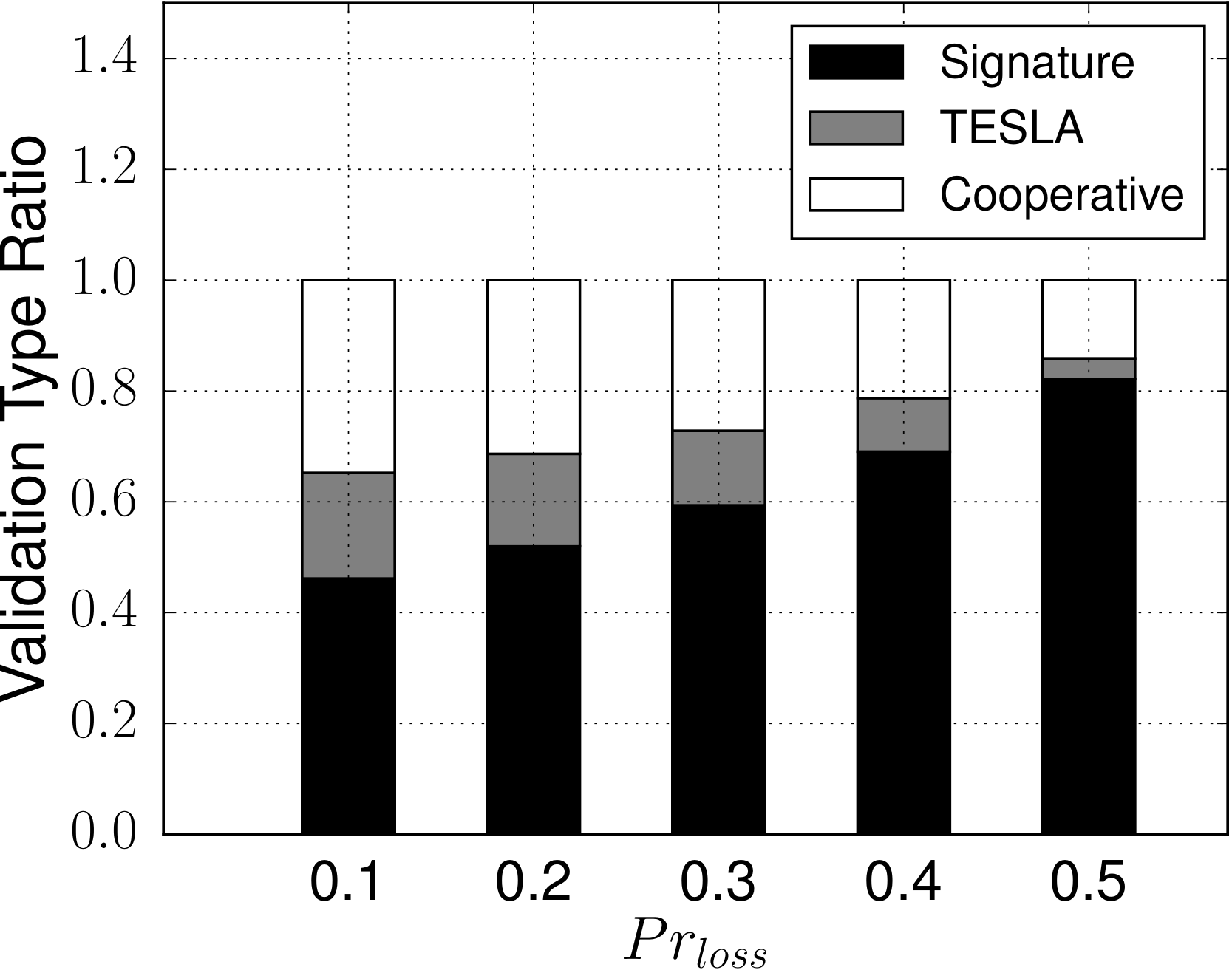}%
		\caption{}%
		\label{subfig_static_type_loss}%
	\end{subfigure}
	\caption{Benign and Static: Ratio of validated beacons based on different validation types as a function of $N$ with (\subref{subfig_static_type_baseline}) baseline scheme ($\alpha = 0$ and TESLA for $N \geq 40$) and (\subref{subfig_static_type_node}) cooperative beacon verification, and as a function of (\subref{subfig_static_type_num}) $\alpha$ and (\subref{subfig_static_type_loss}) $Pr_{loss}$ under default settings.}
	\label{fig_static_type}
\end{figure*}
\begin{figure*}[h!]
	\centering
	\begin{subfigure}{.24\textwidth}
		\includegraphics[width=\columnwidth]{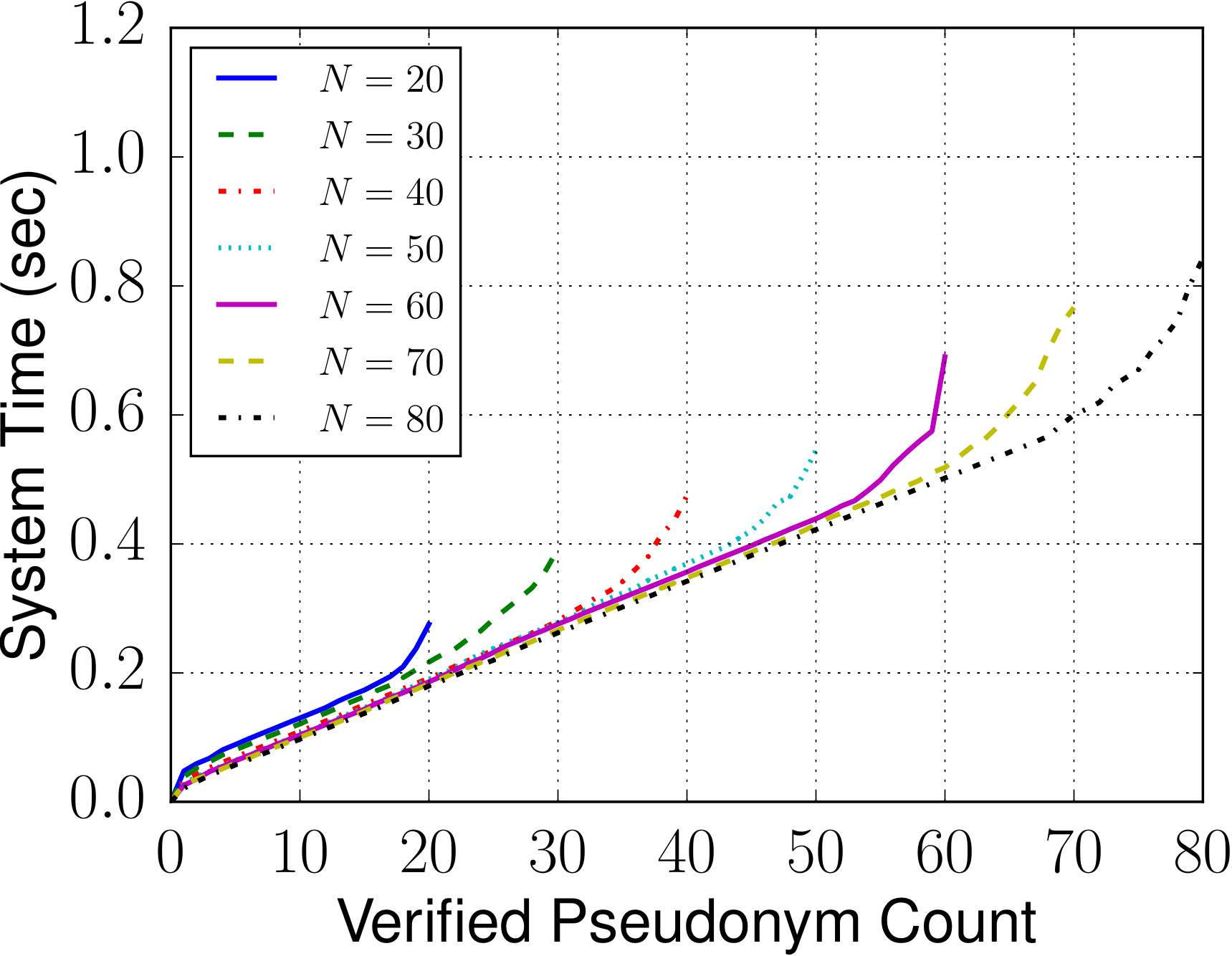}%
		\caption{}%
		\label{subfig_static_psnym_node}%
	\end{subfigure}
	\begin{subfigure}{.24\textwidth}
		\includegraphics[width=\columnwidth]{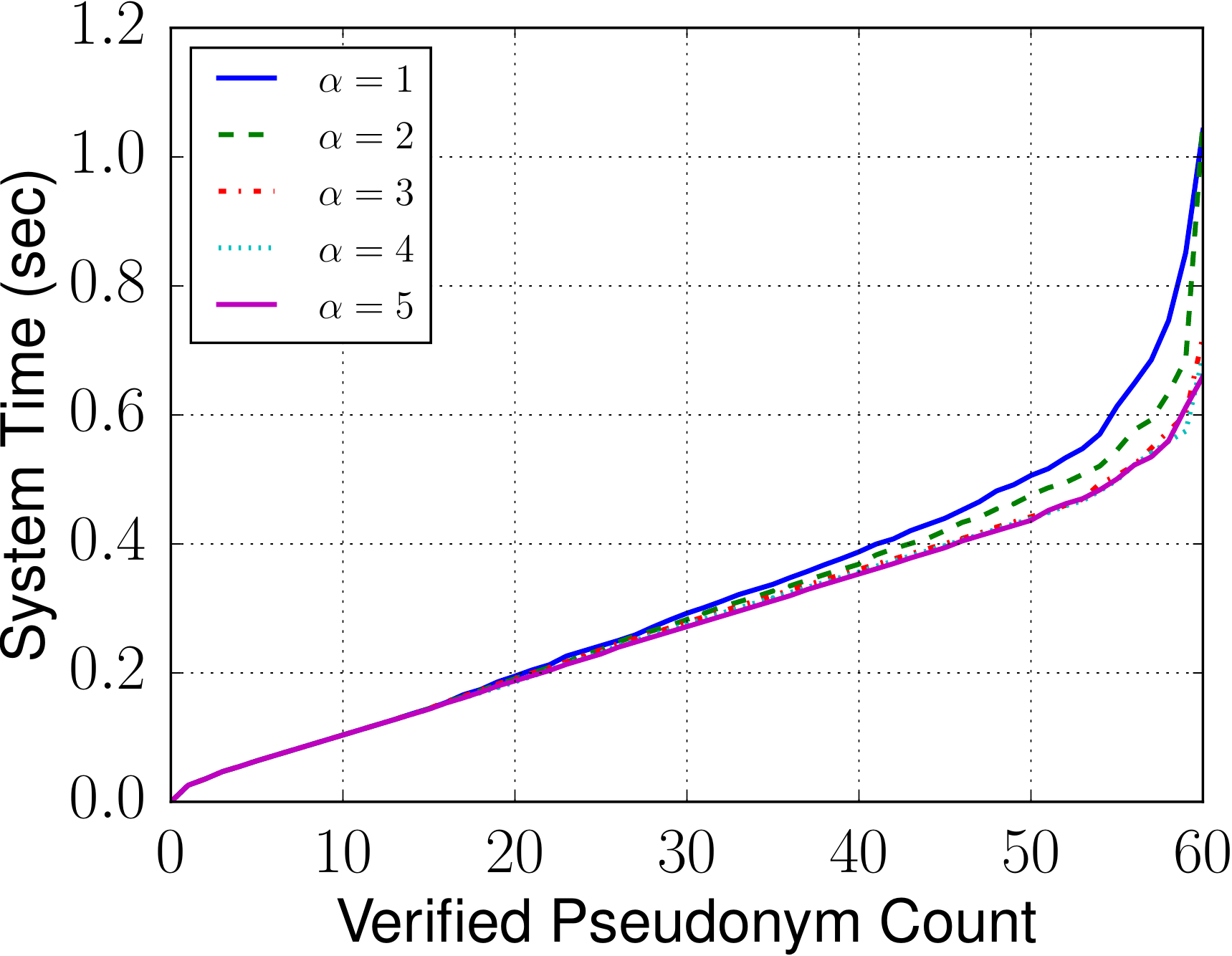}%
		\caption{}%
		\label{subfig_static_psnym_num}%
	\end{subfigure}
	\begin{subfigure}{.24\textwidth}
		\includegraphics[width=\columnwidth]{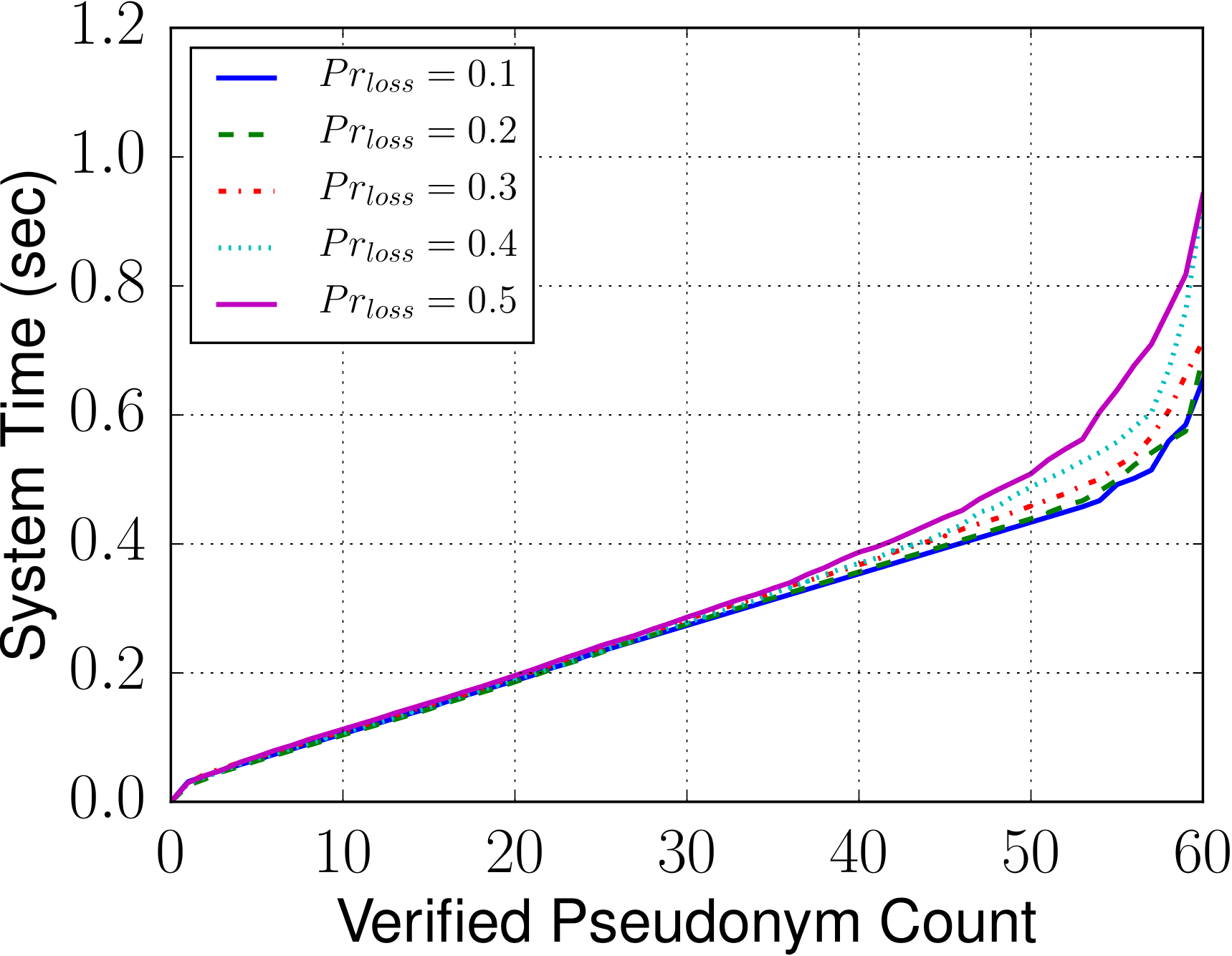}%
		\caption{}%
		\label{subfig_static_psnym_loss}%
	\end{subfigure}
	\caption{Benign and Static: Passed system time as a function of total verified pseudonyms (out of $N$) with different (\subref{subfig_static_psnym_node}) $N$, (\subref{subfig_static_psnym_num}) $\alpha$ and (\subref{subfig_static_psnym_loss}) $Pr_{loss}$.}
	\label{fig_static_psnym}
\end{figure*}
\begin{figure*}[t]
	\centering
	\begin{subfigure}{.24\textwidth}
		\includegraphics[width=\columnwidth]{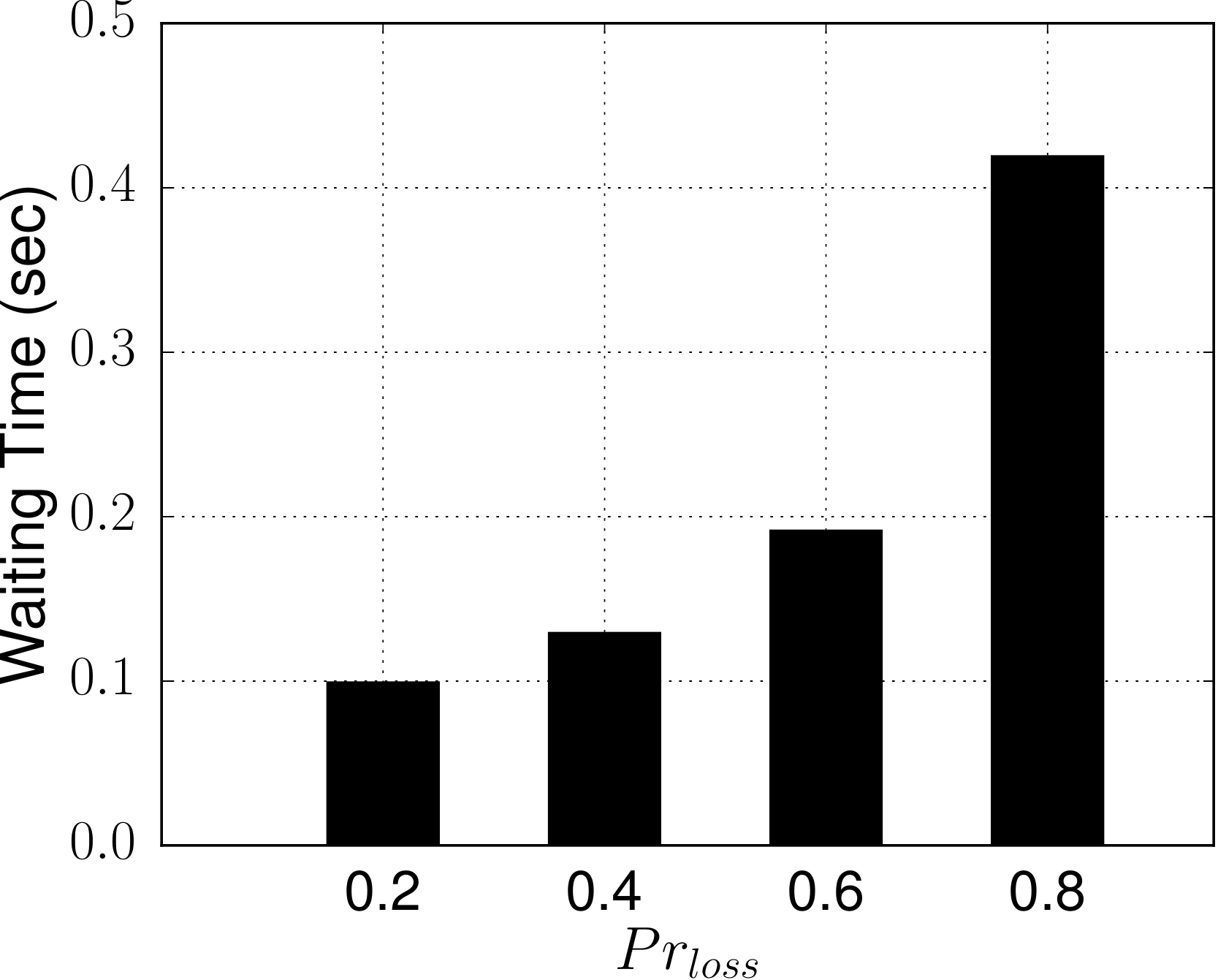}%
		\caption{}%
		\label{subfig_static_adv_waiting_num_2}%
	\end{subfigure}\hspace{1mm}
	\begin{subfigure}{.24\textwidth}
		\includegraphics[width=\columnwidth]{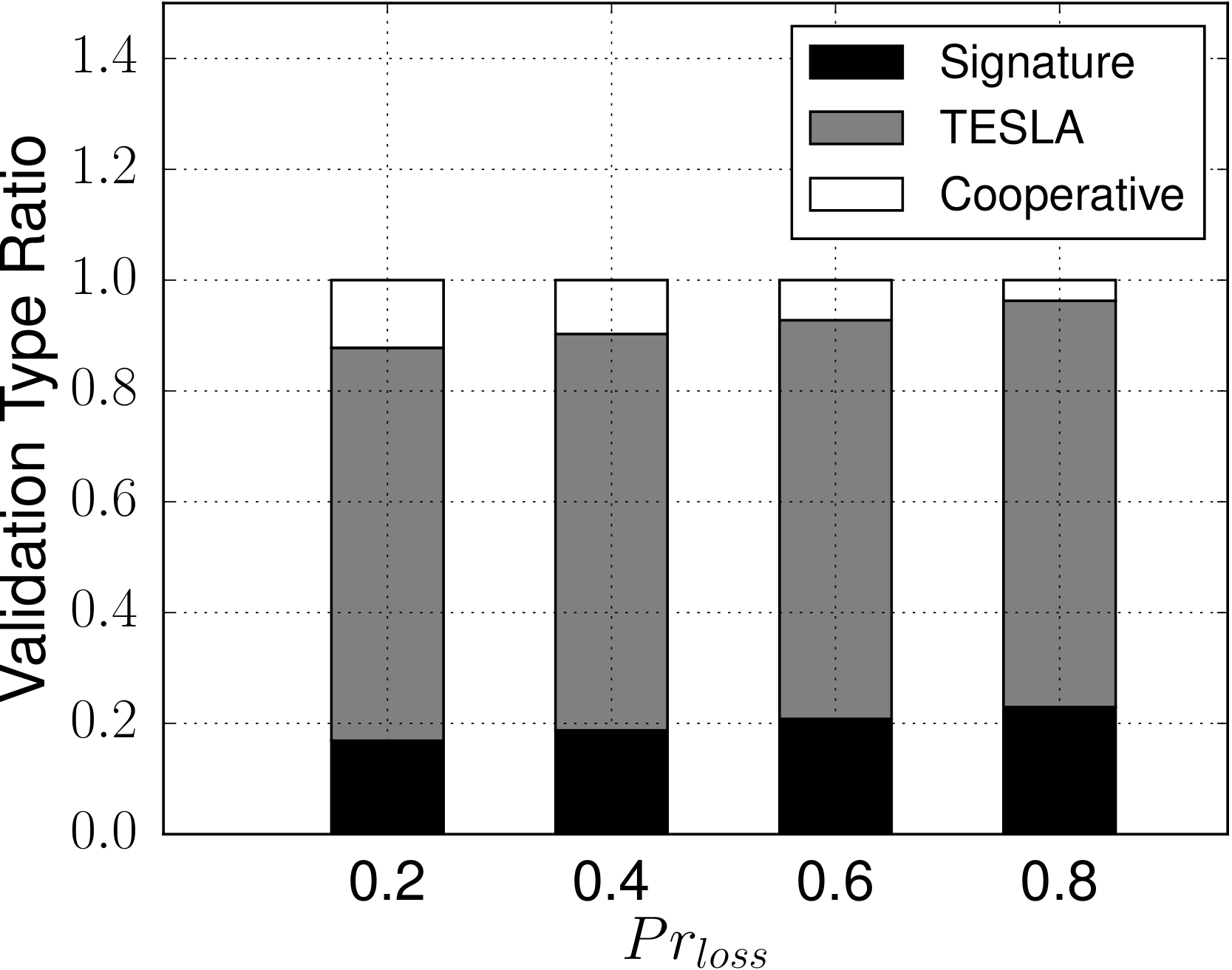}%
		\caption{}%
		\label{subfig_static_adv_type_num_2}%
	\end{subfigure}\hspace{1mm}
	\begin{subfigure}{.24\textwidth}
		\includegraphics[width=\columnwidth]{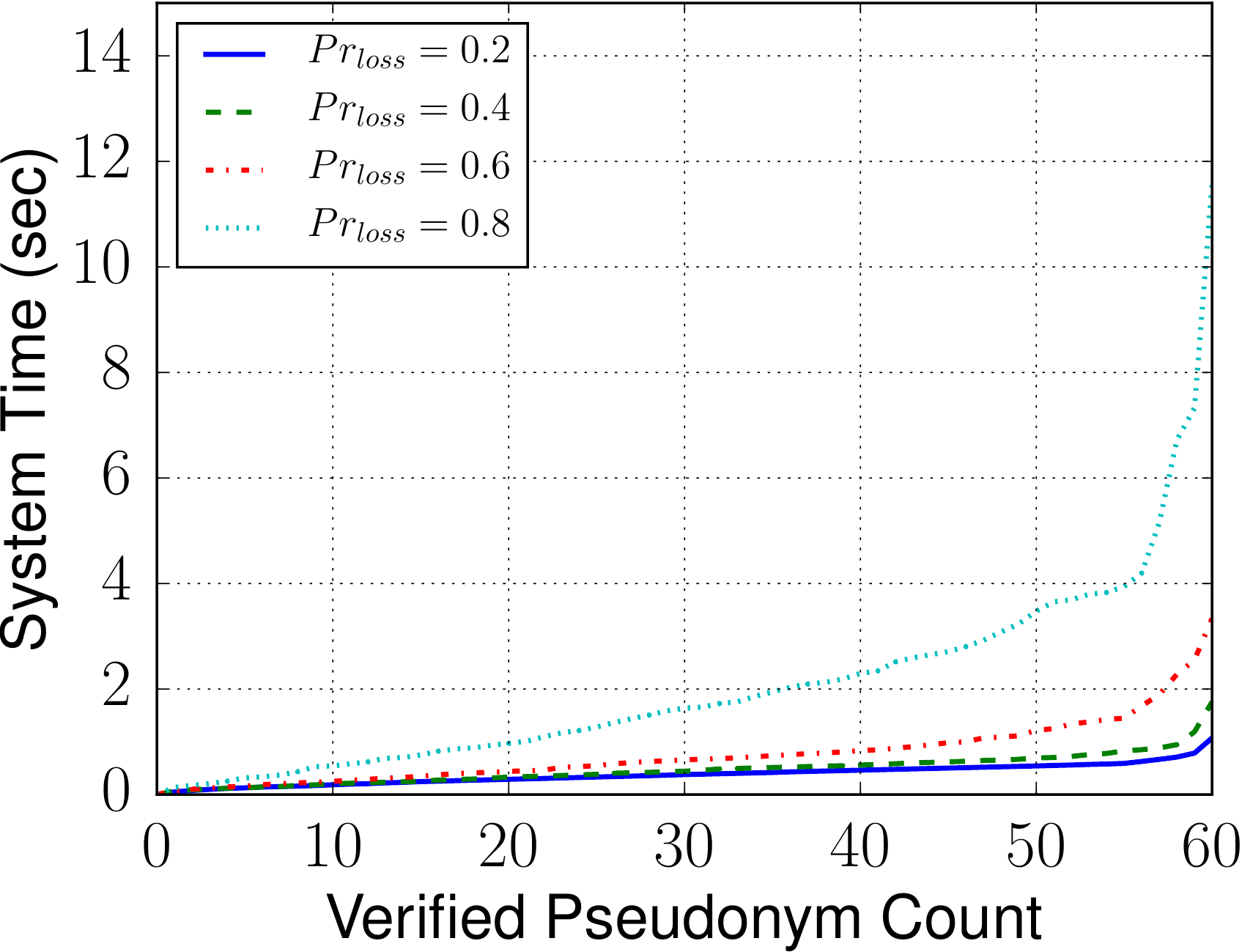}%
		\caption{}%
		\label{subfig_static_adv_psnym_num_2}%
	\end{subfigure}\hspace{1mm}

	\begin{subfigure}{.24\textwidth}
		\includegraphics[width=\columnwidth]{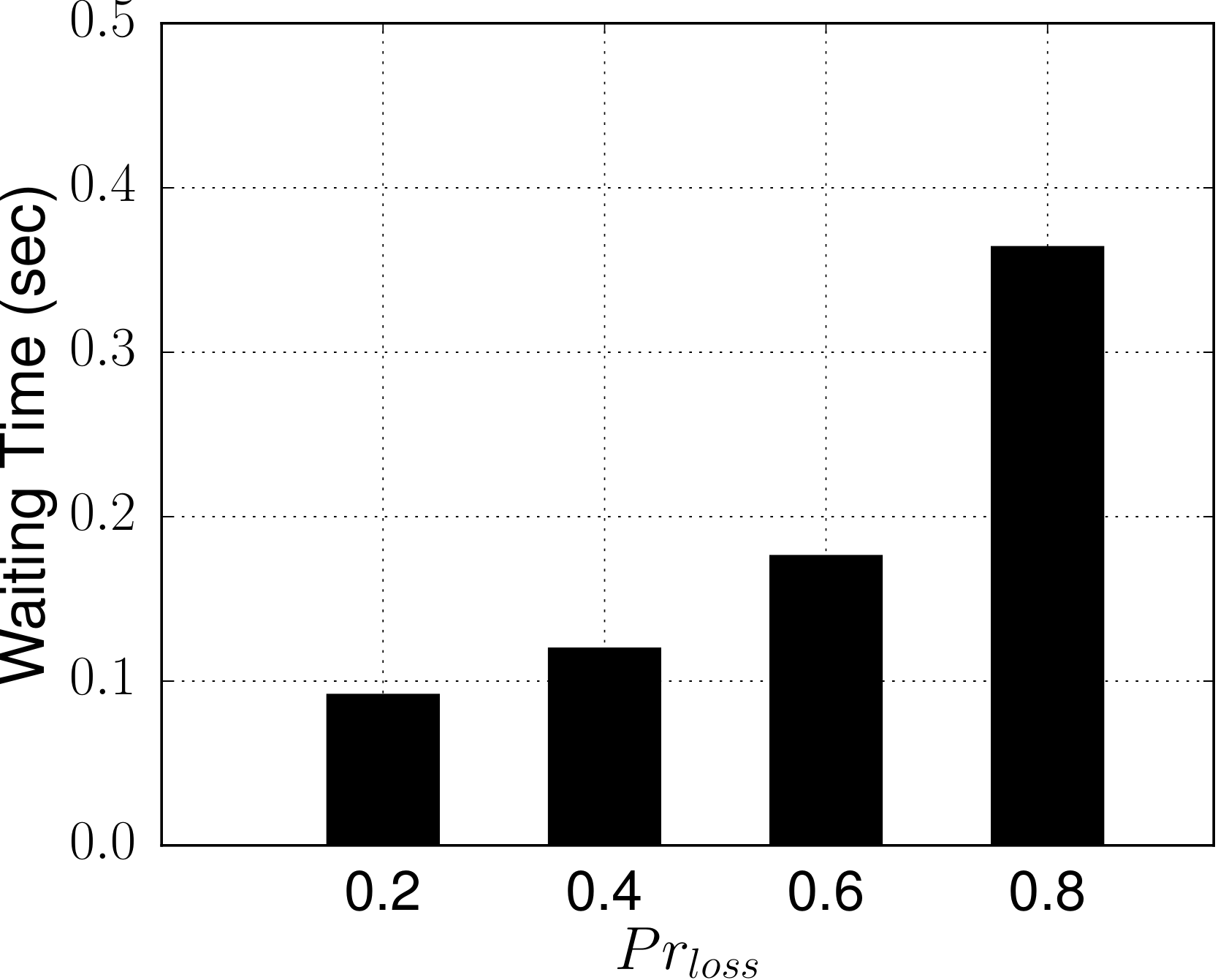}%
		\caption{}%
		\label{subfig_static_adv_waiting_num_4}%
	\end{subfigure}\hspace{1mm}
	\begin{subfigure}{.24\textwidth}
		\includegraphics[width=\columnwidth]{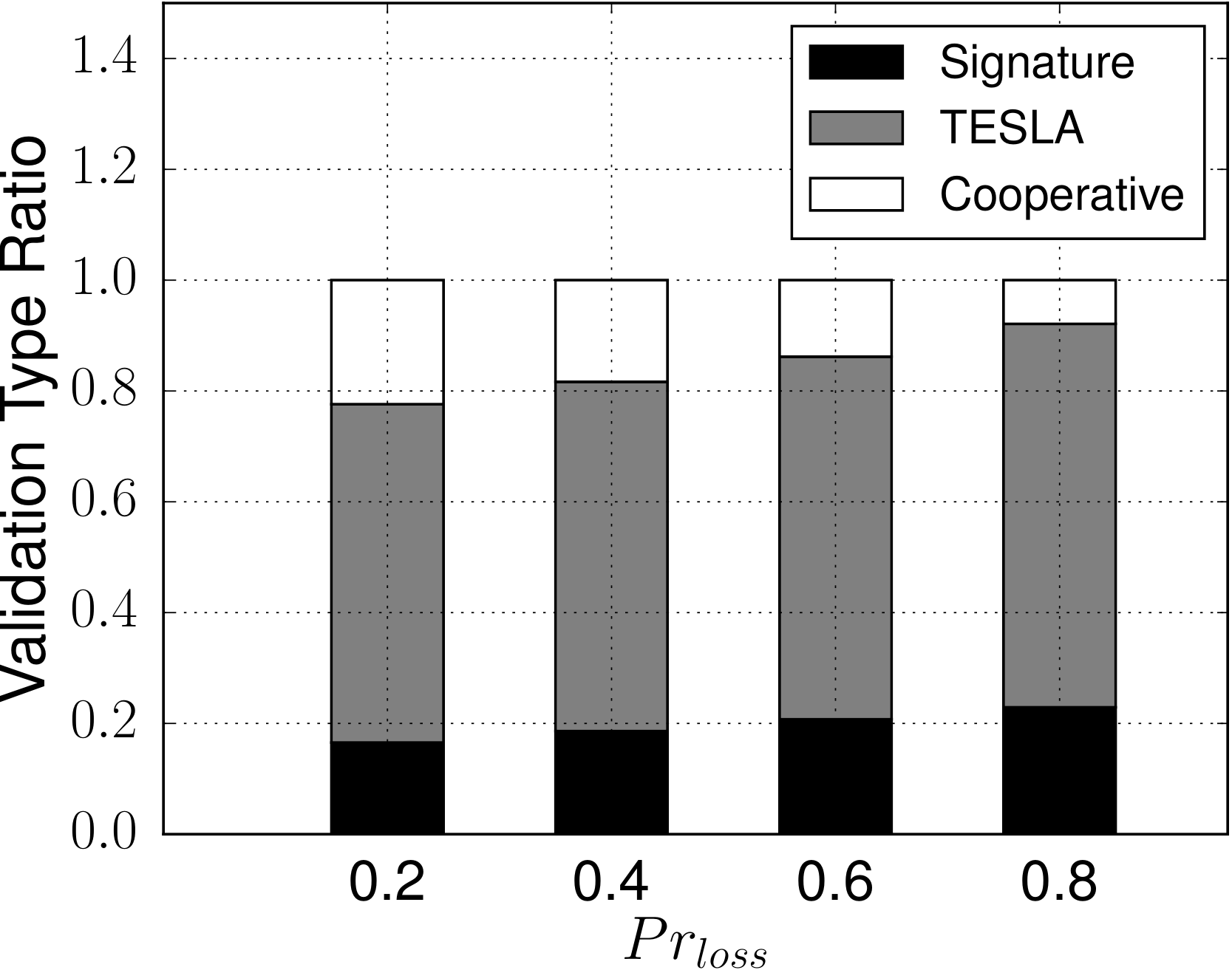}%
		\caption{}%
		\label{subfig_static_adv_type_num_4}%
	\end{subfigure}
	\begin{subfigure}{.24\textwidth}
		\includegraphics[width=\columnwidth]{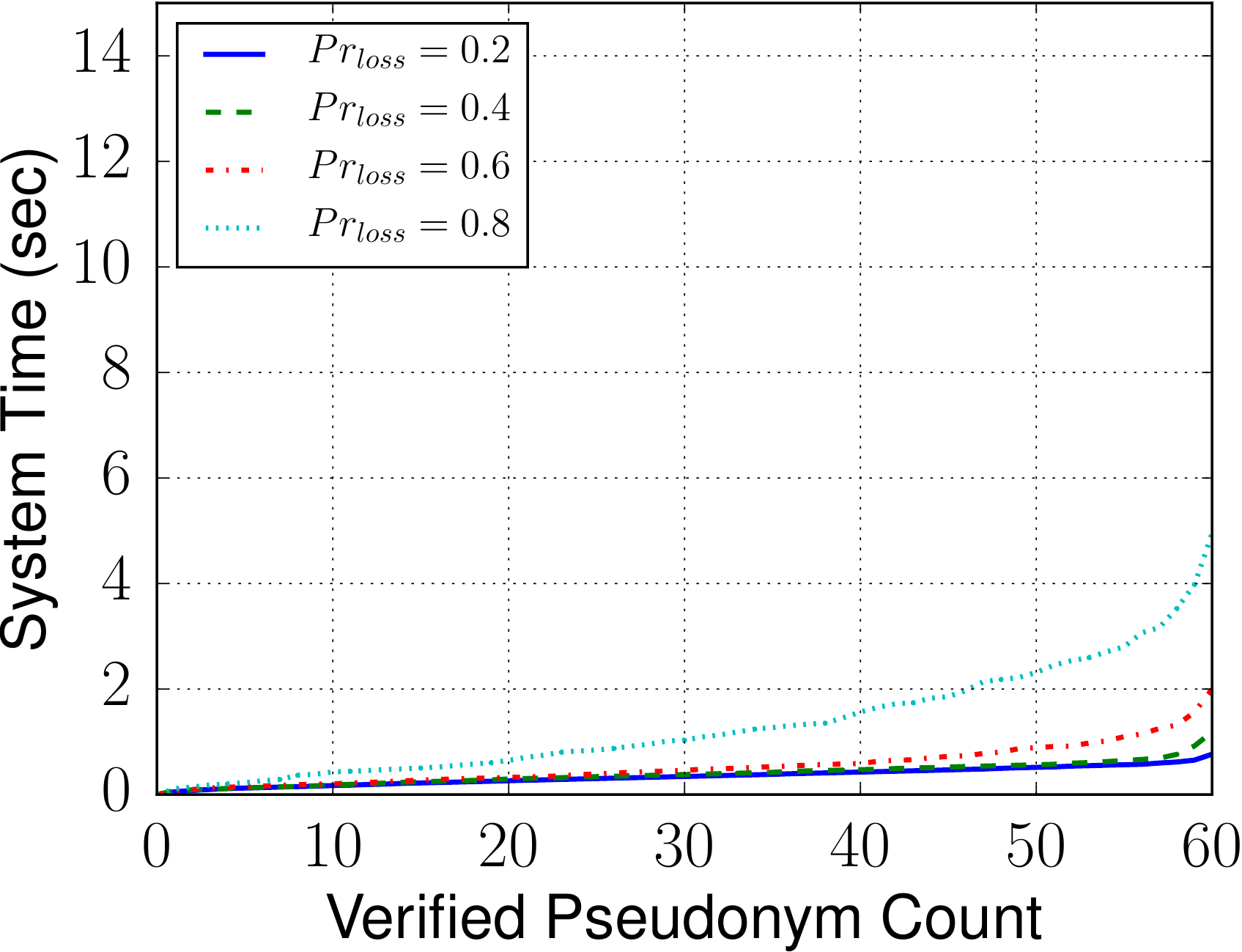}%
		\caption{}%
		\label{subfig_static_adv_psnym_num_4}%
	\end{subfigure}\hspace{1mm}
	\caption{Adversarial and Static: Average waiting time as a function of $Pr_{loss}$ with (\subref{subfig_static_adv_waiting_num_2}) $\alpha = 2$ and (\subref{subfig_static_adv_waiting_num_4}) $\alpha = 4$. Ratio of beacon validation types as a function of $Pr_{loss}$ with (\subref{subfig_static_adv_type_num_2}) $\alpha = 2$ and (\subref{subfig_static_adv_type_num_4}) $\alpha = 4$.}
	\label{fig_static_adv_waiting}
\end{figure*}

\textbf{Benign network:} We start from the evaluation under benign network conditions, without adversarial nodes launching a DDoS attack. All nodes start beaconing at same time (at the beginning of each simulation run). This essentially simulates the situation that vehicles reach a time point for \ac{PC} change, which can be considered as the most challenging situation in a benign network.

\cref{fig_static_waiting} shows average waiting time and \cref{fig_static_type} shows ratio of validated beacons based on different validation types. \emph{Waiting time} of a beacon is defined as the waited time in queue (from the reception time point) before its validation/verification. \cref{subfig_static_waiting_baseline} shows the average waiting time for the baseline scheme as a function of $N$. For $N=20$ and $N=30$, we use baseline scheme without TESLA or cooperative verification (i.e., $\alpha=0$). When $N \geq 40$, the queue would not be sustainable: the average message arrival rate would be $N*\gamma*(1-Pr_{loss}) \geq 320$ $Hz$, while only at most $250$ (i.e., $\frac{1\ s}{T_{vrfc}}$) signature verifications can be performed per second. Therefore, we use TESLA~\cite{hu2006strong,studer2009flexible} without cooperative verification for $N \geq 40$. We see when $N \geq 40$, the average waiting time is around $0.1$ $s$, much higher than that for $N \leq 30$. This is due to a significant ratio of beacon messages need to be validated based on TESLA MACs, as shown in Fig.~\ref{subfig_static_type_baseline}. \cref{subfig_static_waiting_node} shows the average waiting time with our cooperative verification scheme under default settings (see Table~\ref{table:parameter}). The average waiting time is significantly decreased thanks to shared verification results. For example, for the default settings ($N=60$), the average waiting time is decreased from around $0.1$ $s$ (baseline scheme with TESLA) to $0.045$ $s$ (our scheme). From Fig.~\ref{subfig_static_type_node}, we see that, when $N$ is high, a significant ratio of beacons are validated cooperatively or based on TESLA MACs while a smaller ratio of beacons is validated based on signatures.

Fig.~\ref{subfig_static_waiting_num} shows the average waiting time as a function of $\alpha$. The average waiting time decreases with larger $\alpha$ because more beacons can be validated based on shared verification results, as shown in Fig.~\ref{subfig_static_type_num}. However, the improvement becomes moderate when $\alpha$ is higher (e.g., from $\alpha=4$ to $\alpha=5$). Therefore, a reasonable $\alpha$ can be chosen based on different network scenarios to introduce minimum communication overhead (for the attached verification results). The average waiting time also decreases with higher $Pr_{loss}$ (Fig.~\ref{subfig_static_waiting_loss}) because less beacons can be received. As a result, a higher ratio of beacons can be validated based on signature verifications (Fig.~\ref{subfig_static_type_loss}).

\cref{fig_static_psnym} shows the progression of non-cached \ac{PC} verification under different settings. For example, with $N=20$ in \cref{subfig_static_psnym_node}, all \acp{PC} can be verified after around $0.25\ s$ from the beginning of simulation. As described earlier, once a beacon message with non-cached \ac{PC} is verified, then the rest of beacon messages attached with the same \ac{PC} can be validated based on TESLA MACs or shared verification results. From \cref{fig_static_psnym}, we see all the \acp{PC} are verified within $1\ s$ under different settings, which indicates the node can gain awareness of all their neighbors within $1\ s$. Once a beacon message from a newly encountered vehicle is verified, the vehicle can continuously keep track of the corresponding vehicle with cheap TESLA-based validations and cooperative beacon verifications. This is beneficial especially for a high neighbor density scenario or under a DDoS attack (see below).

\textbf{Adversarial network:} We further consider the adversarial network scenarios, where benign nodes are under DDoS attacks. The four adversarial nodes start flooding with fictitious beacons (attached with false signatures and \acp{PC}) from the beginning of the simulation and the benign nodes start beaconing after $10\ c$. The simulation finishes after $1$ $min$ (thus, $70\ s$ in total). Here, we also consider relatively high $Pr_{loss}$ (e.g., $0.6$ and $0.8$), because it is normal to have more packet losses when the network is saturated by the fictitious beacons at a high rate.

\cref{subfig_static_adv_waiting_num_2,subfig_static_adv_waiting_num_4} show the average waiting time as a function of $Pr_{loss}$ when $\alpha = 2$ and $\alpha = 4$ respectively. We see with higher $Pr_{loss}$, the average waiting time increases, because the majority of beacons have to be validated based on TESLA MACs, as shown in \cref{subfig_static_adv_type_num_2,subfig_static_adv_type_num_4}. A significant portion of CPU cycles are occupied for verifying fictitious beacons (in an attempt to find new authentic/valid \acp{PC}). For example, when $Pr_{loss} = 0.8$, the average waiting time is more than $0.4\ s$. However, this is still a significant improvement from the baseline scheme, with which the queue would not be even sustainable and waiting time would continuously increase as the time progresses. For example, from simulation results (not shown in the figures), we find that when $\alpha=0$, only a few pseudonyms can be verified within $70\ s$, because the queue is filled with fictitious beacons. This cannot be solved even by setting a lifetime, e.g., $1\ s$, for each received beacon (a beacon is dropped if it stayed in the queue for more than $1\ s$), because the majority of authentic beacons are also dropped due to expiration. We see with higher $\alpha$, the average waiting time slightly decreases (from \cref{subfig_static_adv_waiting_num_2} to \cref{subfig_static_adv_waiting_num_4}) thanks to more beacons can be validated based on shared verification results (\cref{subfig_static_adv_type_num_2,subfig_static_adv_type_num_4}).

From \cref{subfig_static_adv_psnym_num_2,subfig_static_adv_psnym_num_4}, we see that all \acp{PC} can be verified within $3\ s$ when $Pr_{loss} \leq 0.6$. When $Pr_{loss}=0.8$, more time is needed to gain awareness of all neighboring nodes: more than $10\ s$ when $\alpha=2$ and around $5\ s$ when $\alpha=4$. However, as mentioned earlier, this is still a significant improvement considering only a few \acp{PC} can be verified within $60\ s$ after beaconing begins without cooperative verification.

{\color{blue}

\begin{figure*}[t!]
	\centering
	\includegraphics[width=0.9\textwidth]{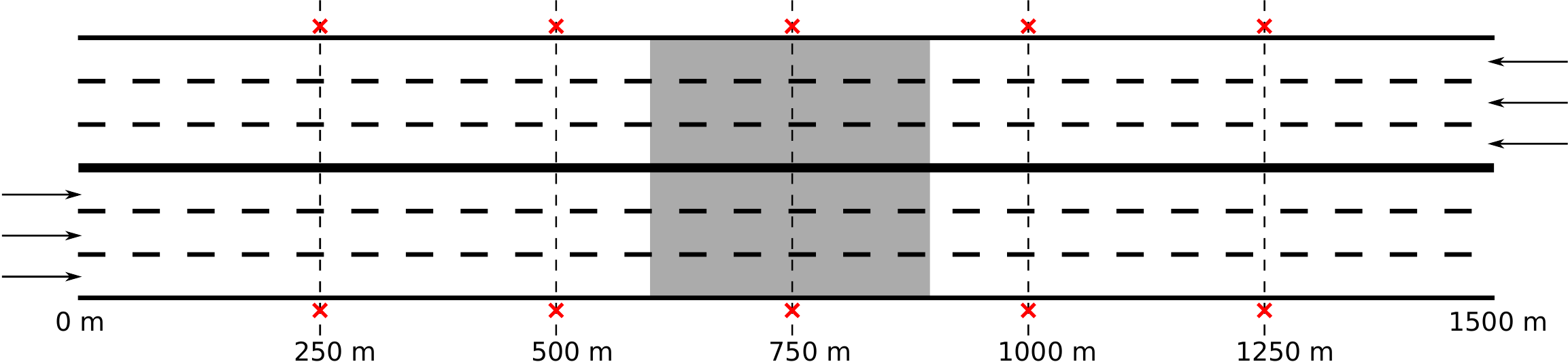}
	\caption{Highway scenario: A six-lane highway scenario with a neighbor density $N \approx 60$; results collected from nodes passing the gray area (central $300\ m$ area). X marks indicate placement of adversarial nodes.}
	\label{fig_mobile}
\end{figure*}

\begin{figure*}[h]
	\centering
	\begin{subfigure}{.24\textwidth}
		\includegraphics[width=\columnwidth]{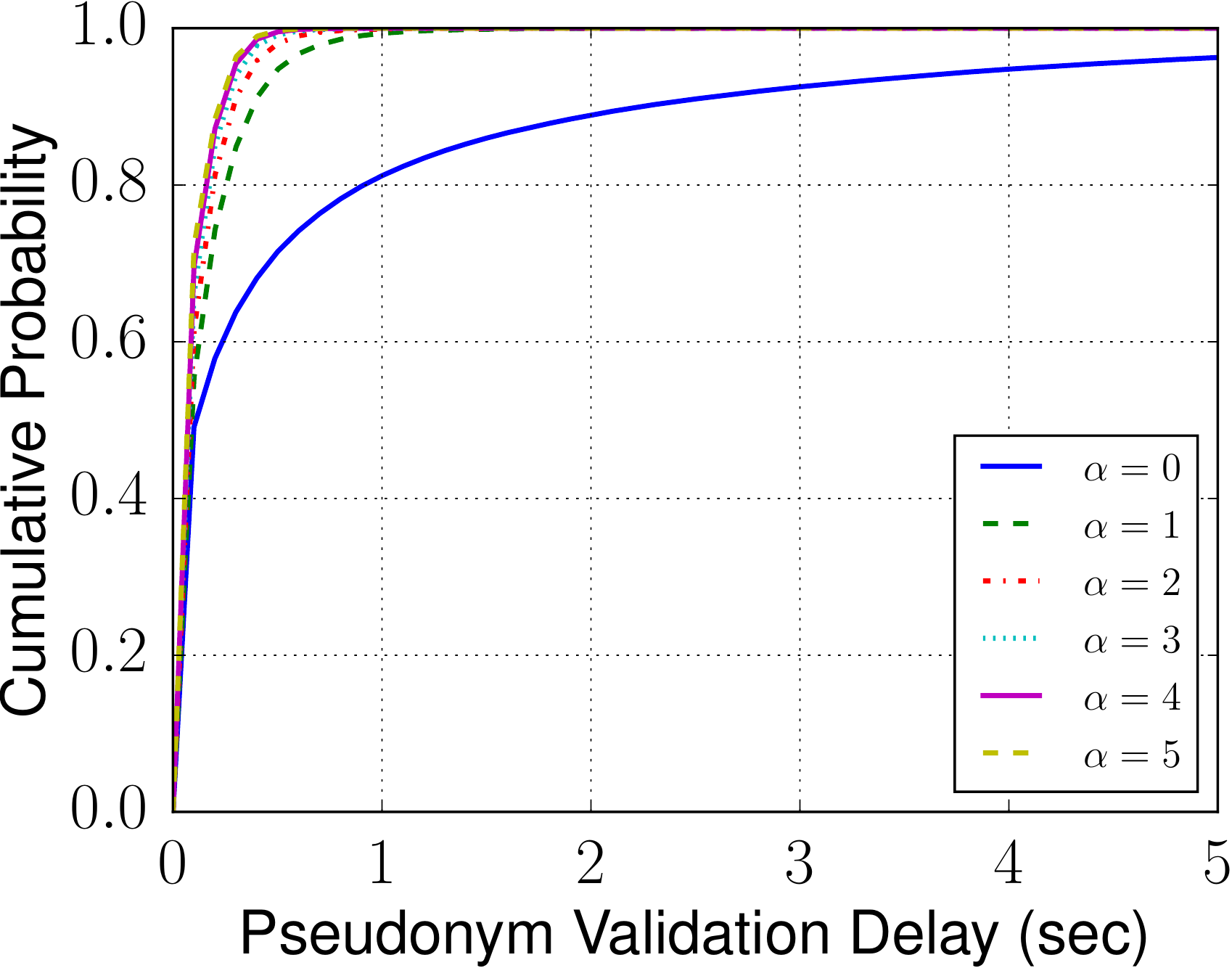}%
		\caption{}%
		\label{subfig_mobile_num_psnym}%
	\end{subfigure}\hspace{1mm}
	\begin{subfigure}{.24\textwidth}
		\includegraphics[width=\columnwidth]{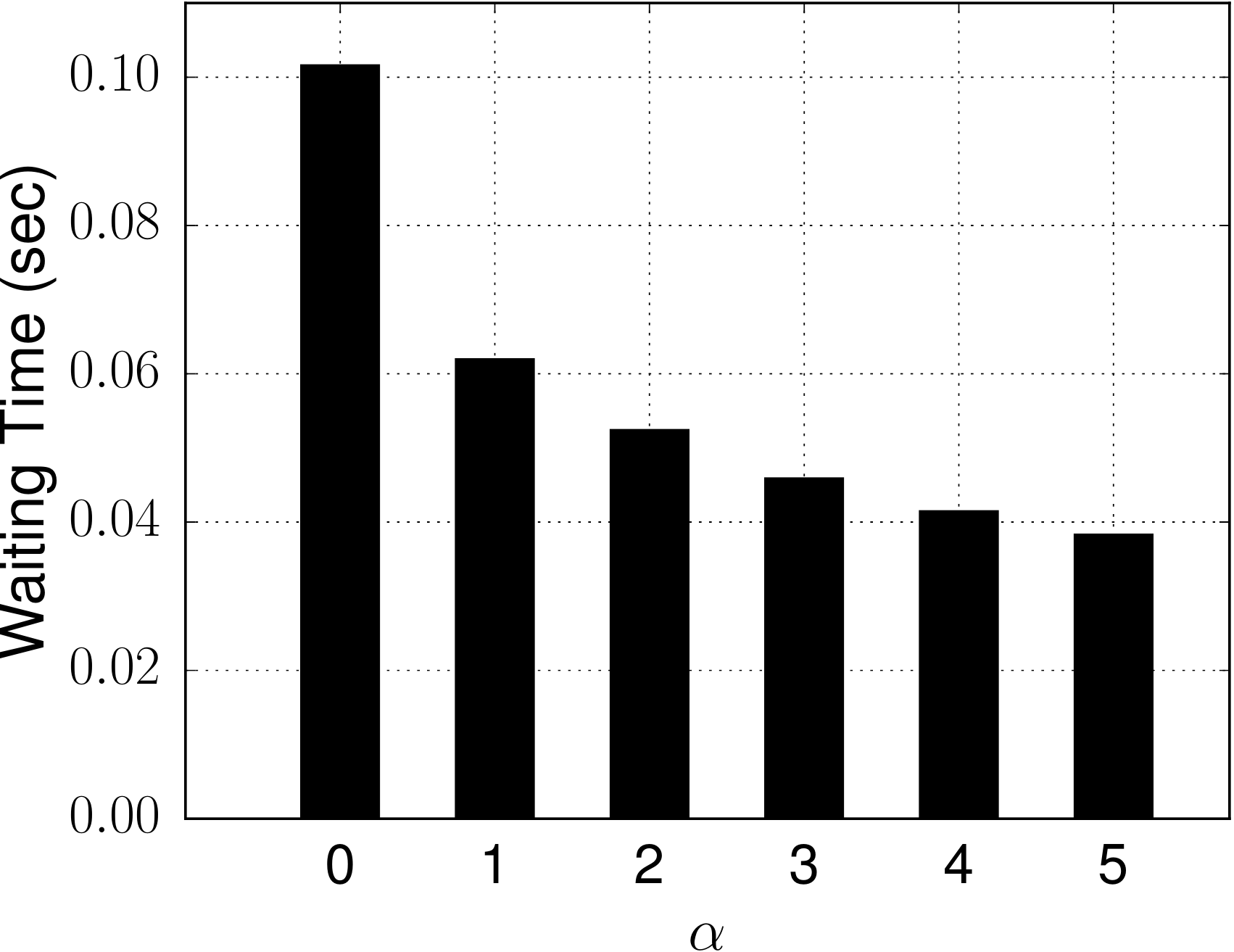}%
		\caption{}%
		\label{subfig_mobile_num_waiting}%
	\end{subfigure}\hspace{1mm}
	\begin{subfigure}{.24\textwidth}
		\includegraphics[width=\columnwidth]{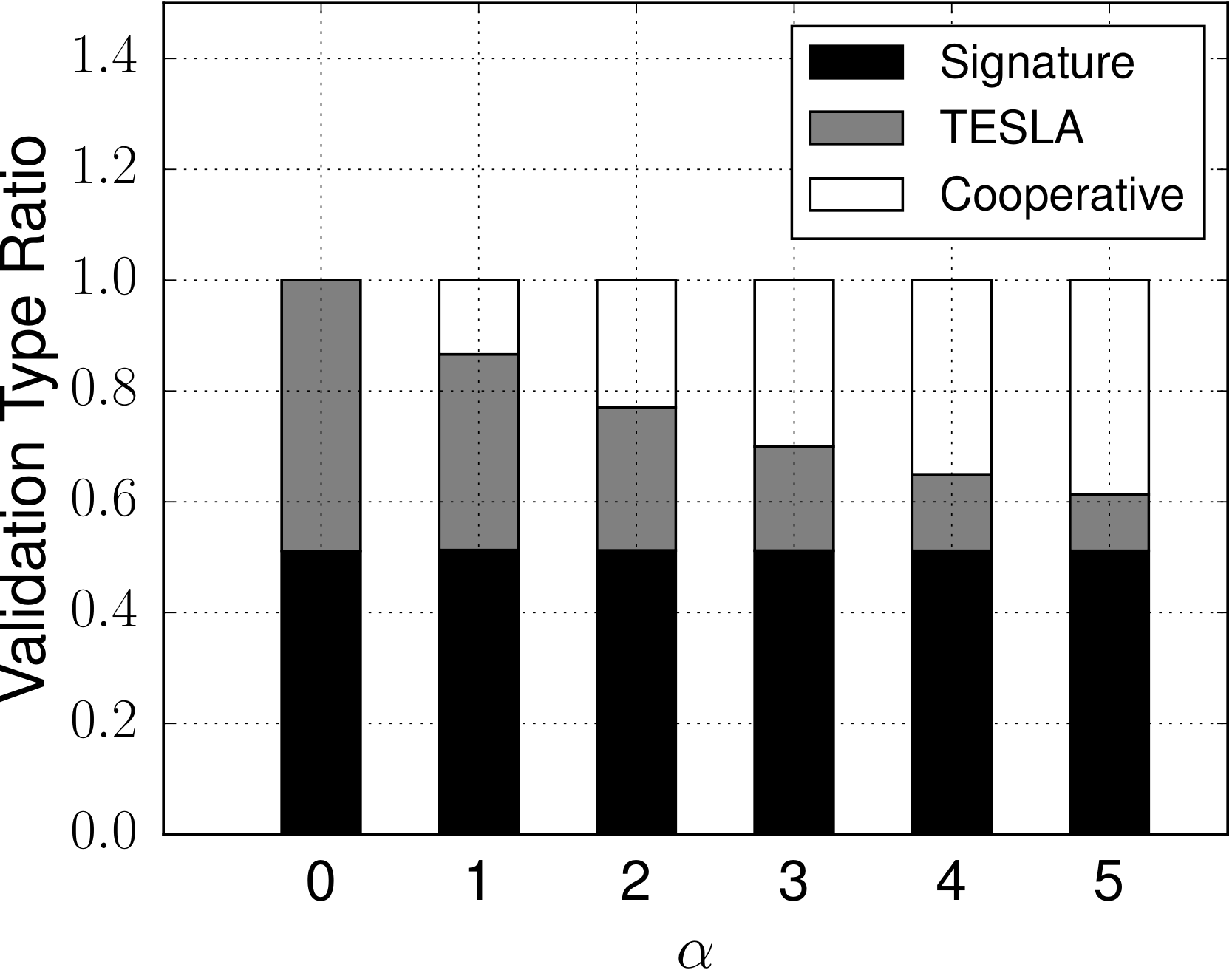}%
		\caption{}%
		\label{subfig_mobile_num_type}%
	\end{subfigure}\hspace{1mm}

	\begin{subfigure}{.24\textwidth}
		\includegraphics[width=\columnwidth]{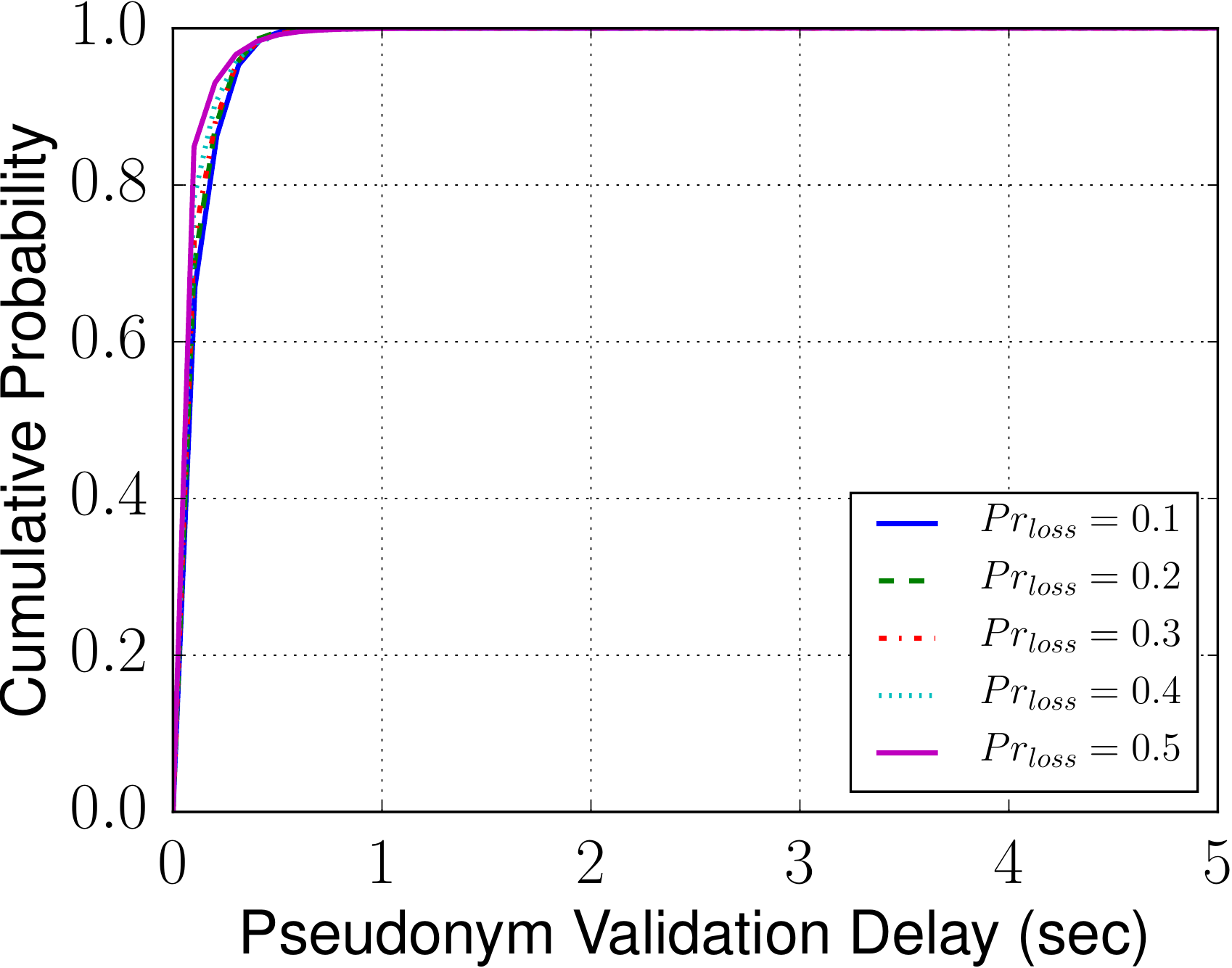}%
		\caption{}%
		\label{subfig_mobile_loss_psnym}%
	\end{subfigure}\hspace{1mm}
	\begin{subfigure}{.24\textwidth}
		\includegraphics[width=\columnwidth]{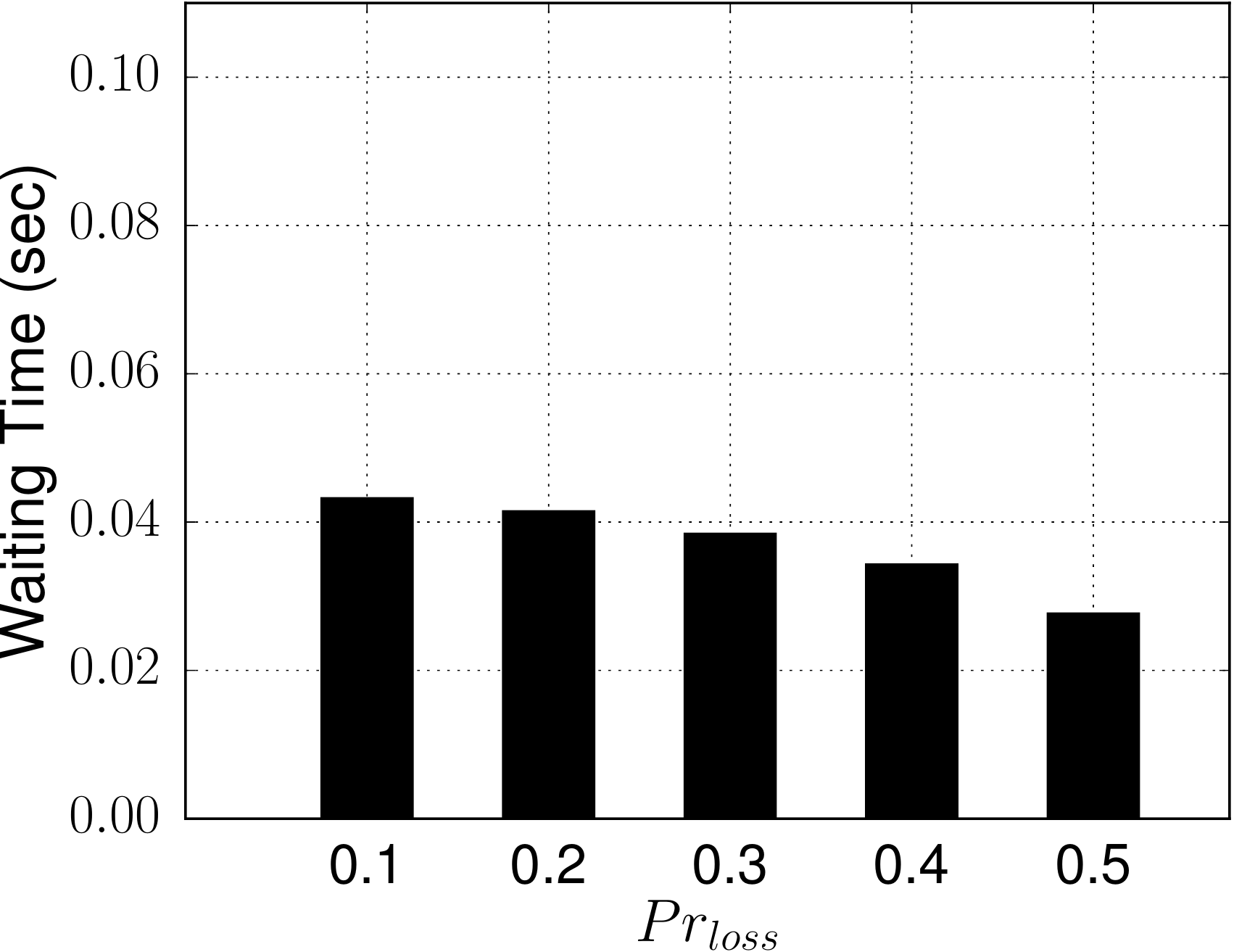}%
		\caption{}%
		\label{subfig_mobile_loss_waiting}%
	\end{subfigure}\hspace{1mm}
	\begin{subfigure}{.24\textwidth}
		\includegraphics[width=\columnwidth]{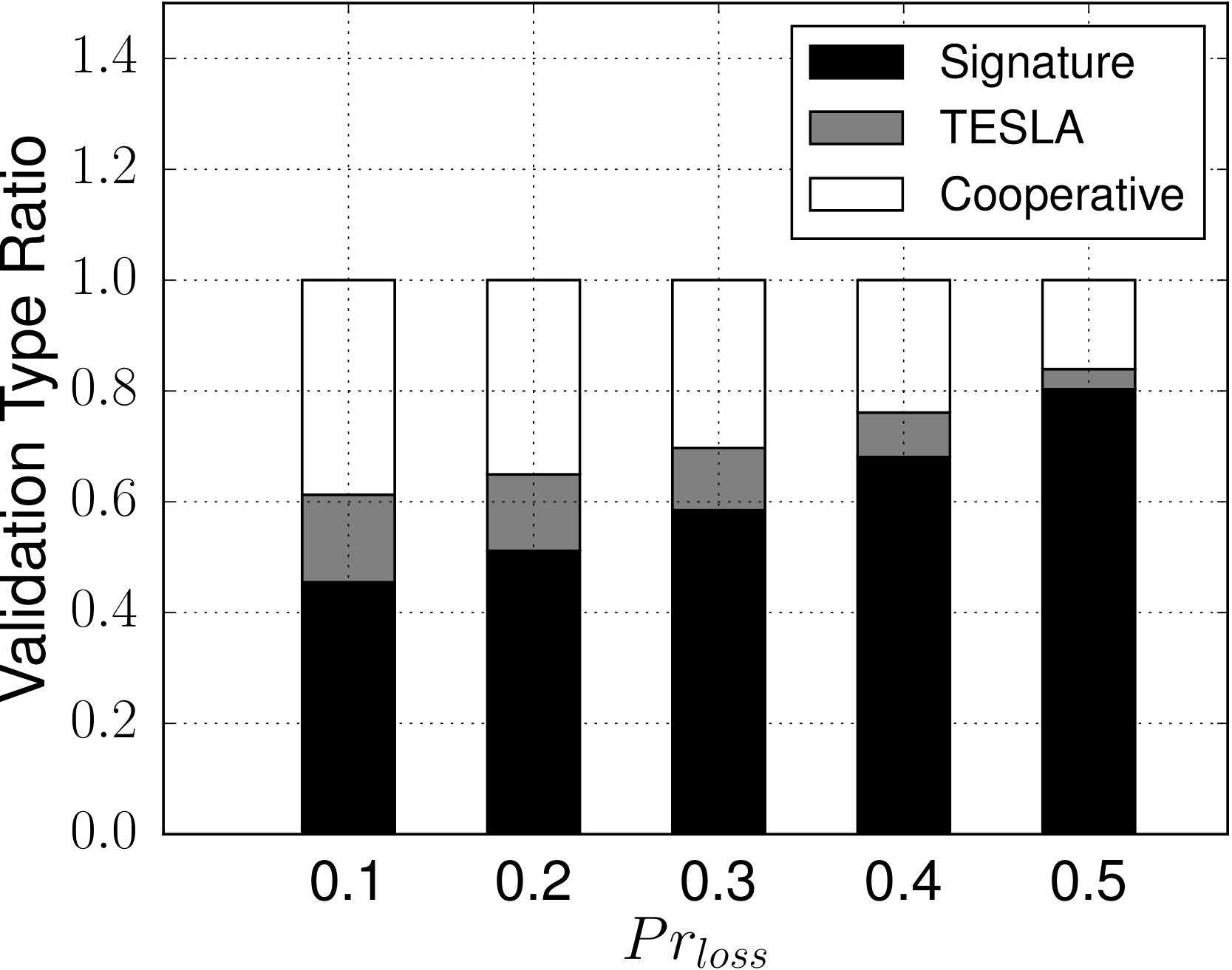}%
		\caption{}%
		\label{subfig_mobile_loss_type}%
	\end{subfigure}
	\caption{Benign and Highway: CDF of pseudonym validation delay as a function of (\subref{subfig_mobile_num_psnym}) $\alpha$ and (\subref{subfig_mobile_loss_psnym}) $Pr_{loss}$. Average waiting time as a function of (\subref{subfig_mobile_num_waiting}) $\alpha$ and (\subref{subfig_mobile_loss_waiting}) $Pr_{loss}$. Ratio of validated beacons based on different validation types as a function of (\subref{subfig_mobile_num_type}) $\alpha$ and (\subref{subfig_mobile_loss_type}) $Pr_{loss}$.}
	\label{fig_mobile_waiting}
\end{figure*}

\begin{figure*}[t]
	\centering
	\begin{subfigure}{.24\textwidth}
		\includegraphics[width=\columnwidth]{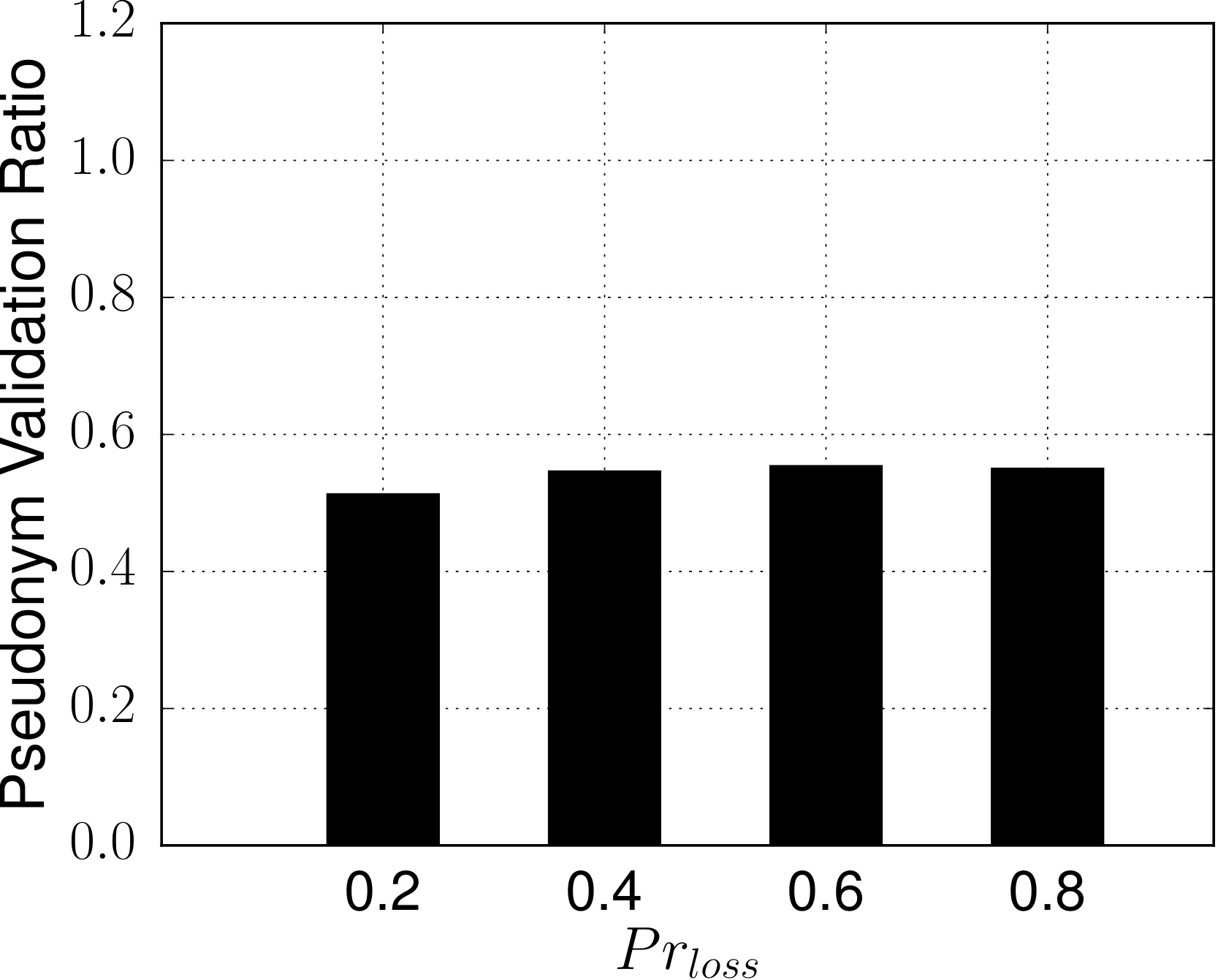}%
		\caption{}%
		\label{subfig_mobile_adv_num_0_ratio}%
	\end{subfigure}\hspace{1mm}
	\begin{subfigure}{.24\textwidth}
		\includegraphics[width=\columnwidth]{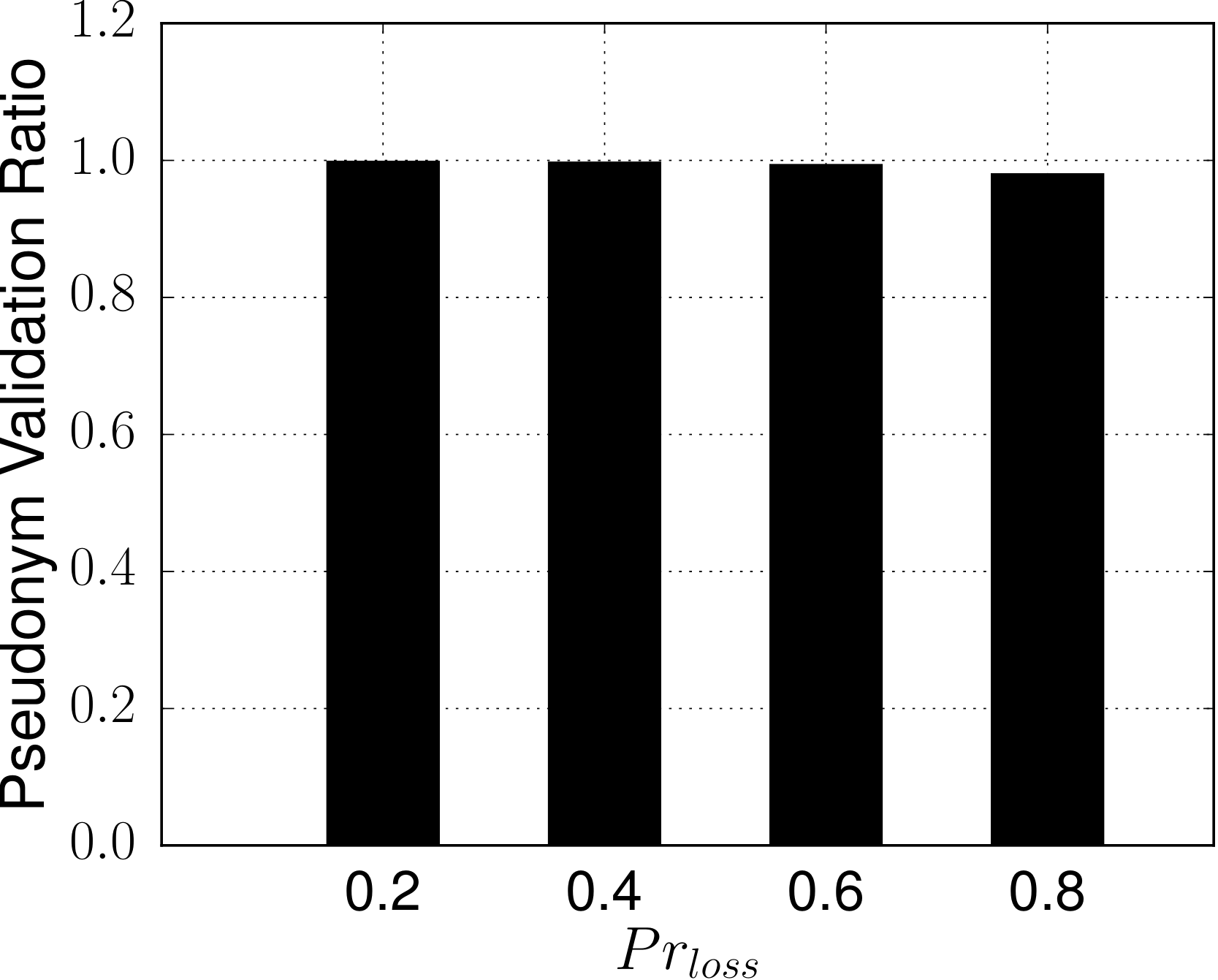}%
		\caption{}%
		\label{subfig_mobile_adv_num_2_ratio}%
	\end{subfigure}\hspace{1mm}
	\begin{subfigure}{.24\textwidth}
		\includegraphics[width=\columnwidth]{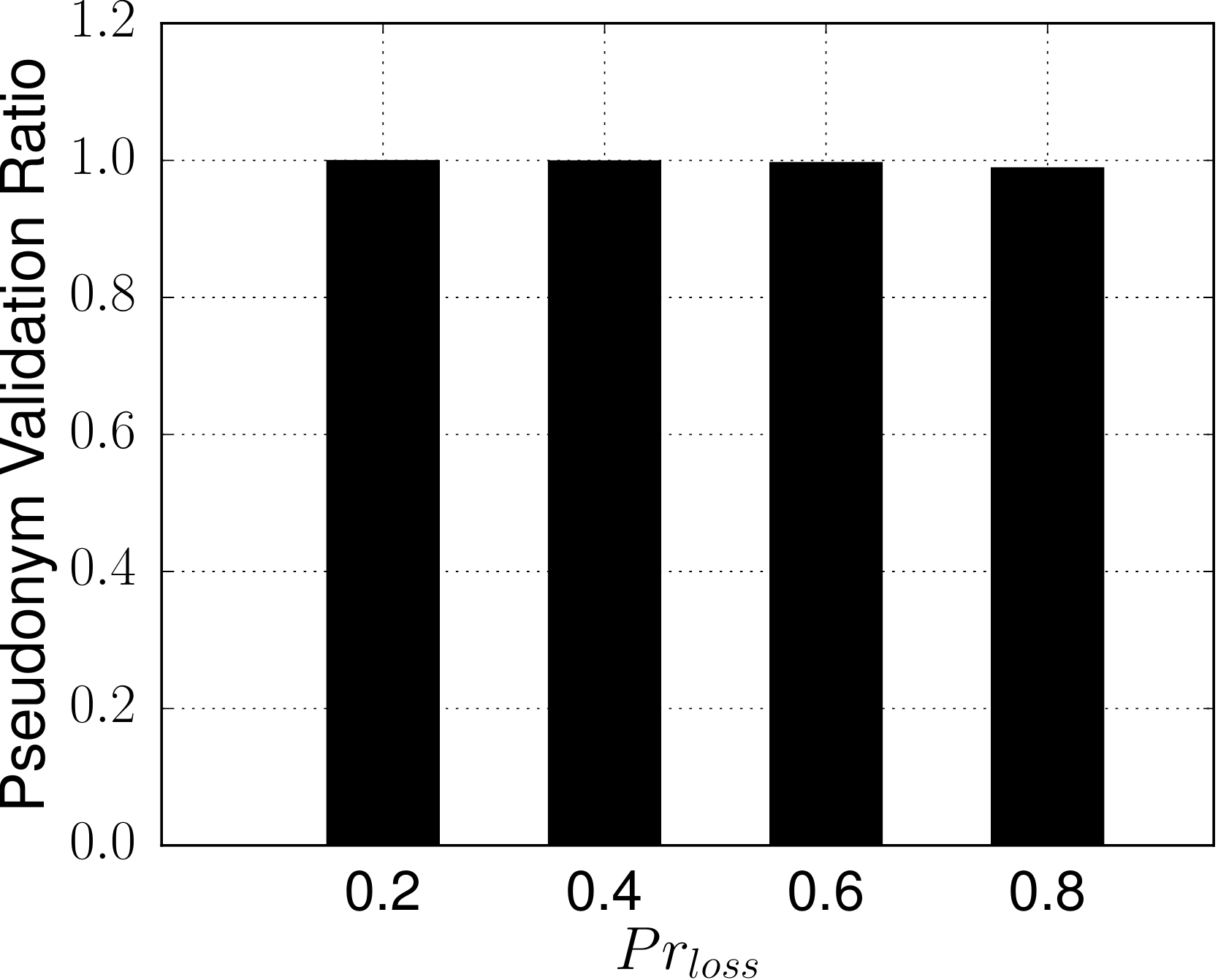}%
		\caption{}%
		\label{subfig_mobile_adv_num_4_ratio}%
	\end{subfigure}\hspace{1mm}
	\caption{Adversarial and Highway: Pseudonym validation ratio as a function of $Pr_{loss}$  with (\subref{subfig_mobile_adv_num_0_ratio}) $\alpha = 0$, (\subref{subfig_mobile_adv_num_2_ratio}) $\alpha = 2$ and (\subref{subfig_mobile_adv_num_4_ratio}) $\alpha = 4$.}
	\label{fig_mobile_adv_ratio}
\end{figure*}

\begin{figure*}[h]
	\centering
	\begin{subfigure}{.24\textwidth}
		\includegraphics[width=\columnwidth]{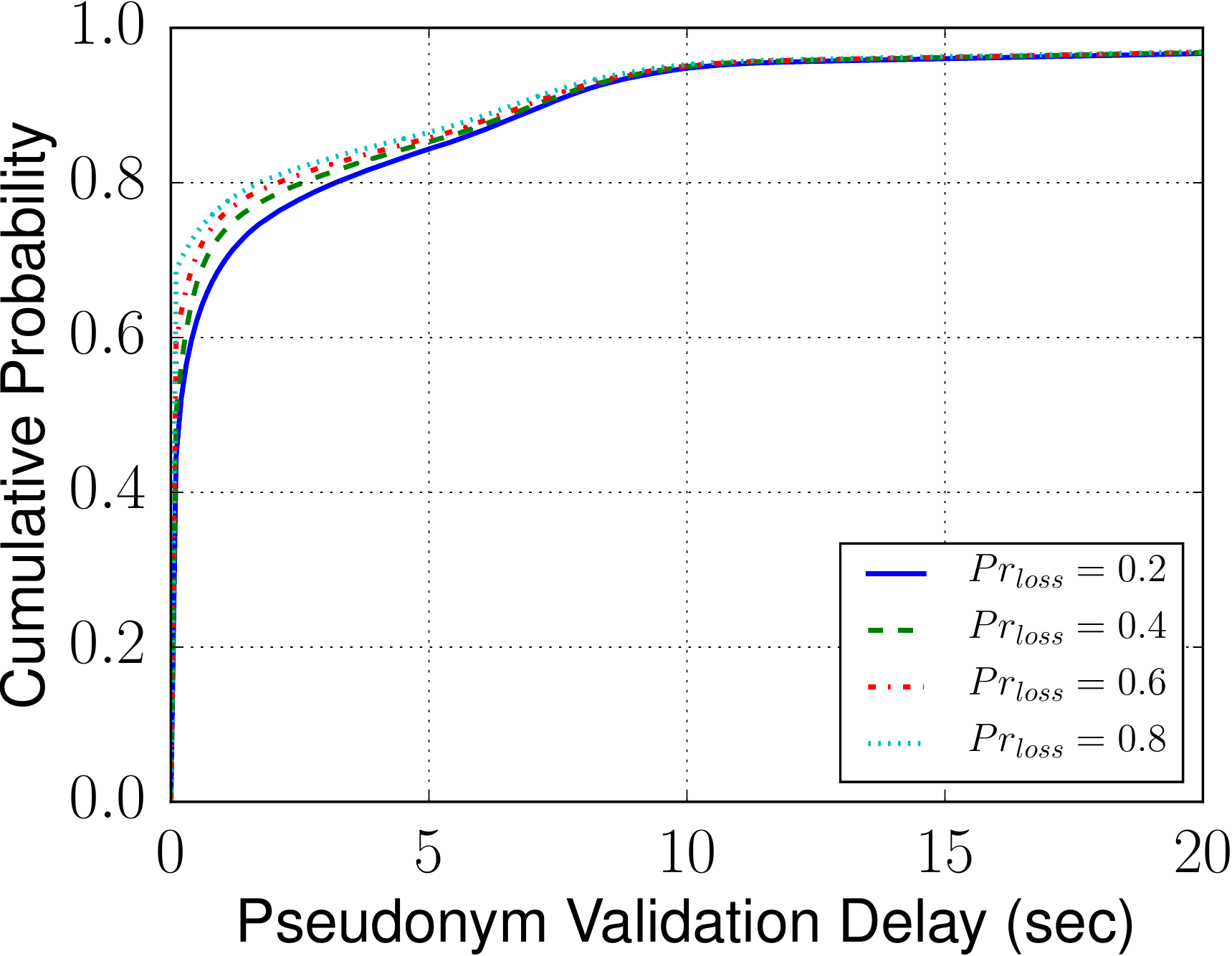}%
		\caption{}%
		\label{subfig_mobile_adv_num_0_psnym}%
	\end{subfigure}\hspace{1mm}
	\begin{subfigure}{.24\textwidth}
		\includegraphics[width=\columnwidth]{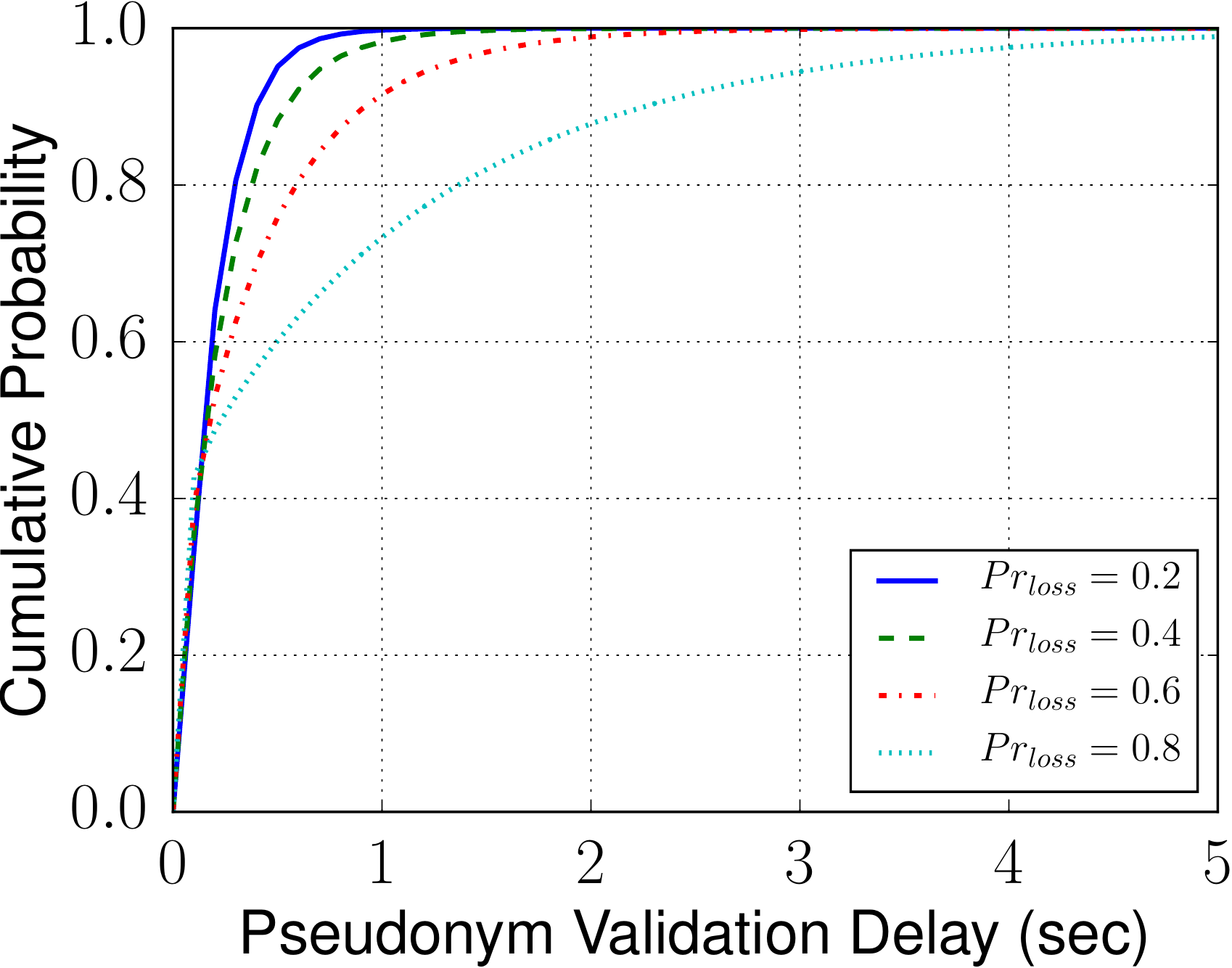}%
		\caption{}%
		\label{subfig_mobile_adv_num_2_psnym}%
	\end{subfigure}\hspace{1mm}
	\begin{subfigure}{.24\textwidth}
		\includegraphics[width=\columnwidth]{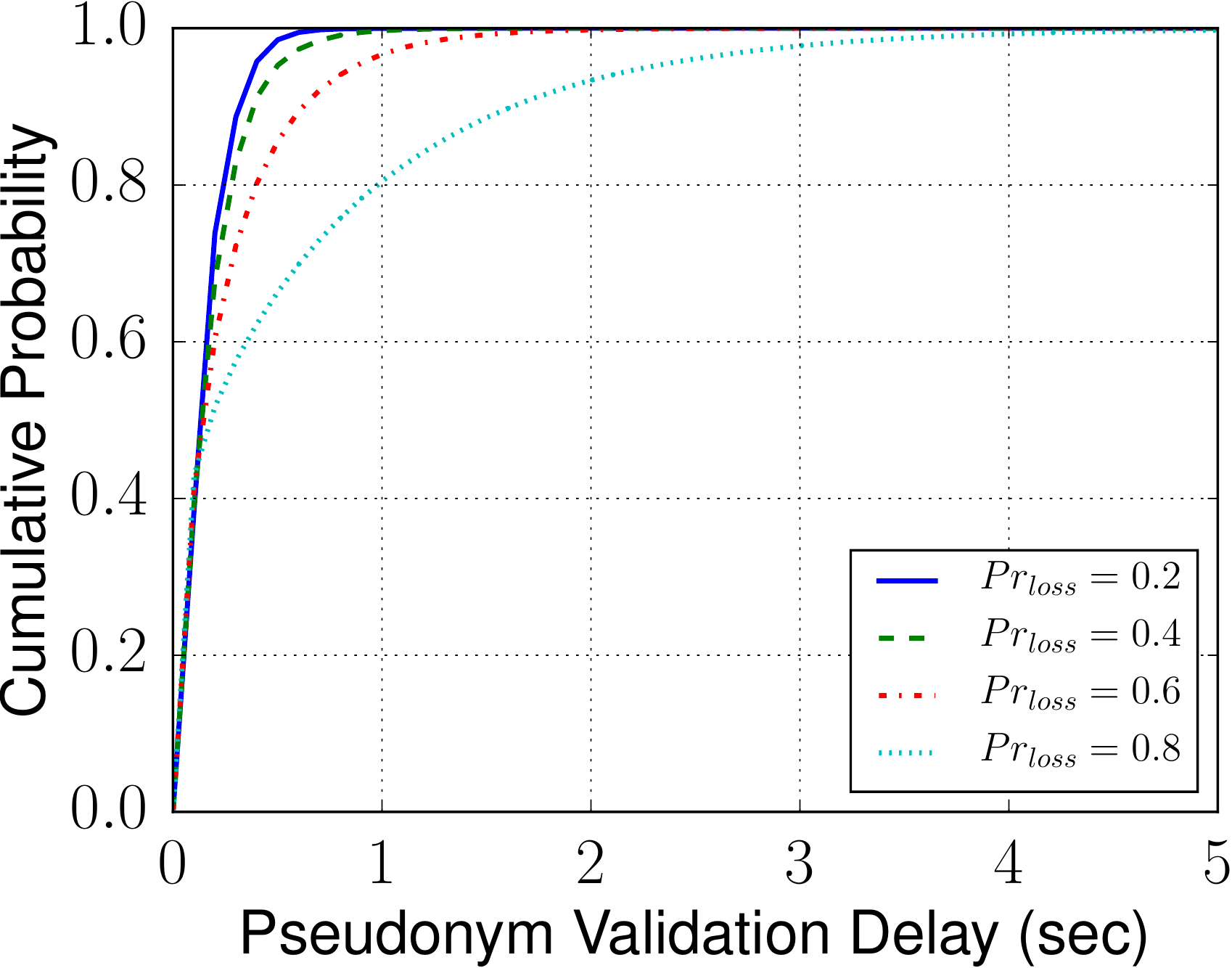}%
		\caption{}%
		\label{subfig_mobile_adv_num_4_psnym}%
	\end{subfigure}\hspace{1mm}
	\caption{Adversarial and Highway: CDF of pseudonym validation delay as a function of $Pr_{loss}$ with (\subref{subfig_mobile_adv_num_0_psnym}) $\alpha = 0$, (\subref{subfig_mobile_adv_num_2_psnym}) $\alpha = 2$ and (\subref{subfig_mobile_adv_num_4_psnym}) $\alpha = 4$.}
	\label{fig_mobile_adv_psnym}
\end{figure*}

\begin{figure*}[h!]
	\centering
	\begin{subfigure}{.24\textwidth}
		\includegraphics[width=\columnwidth]{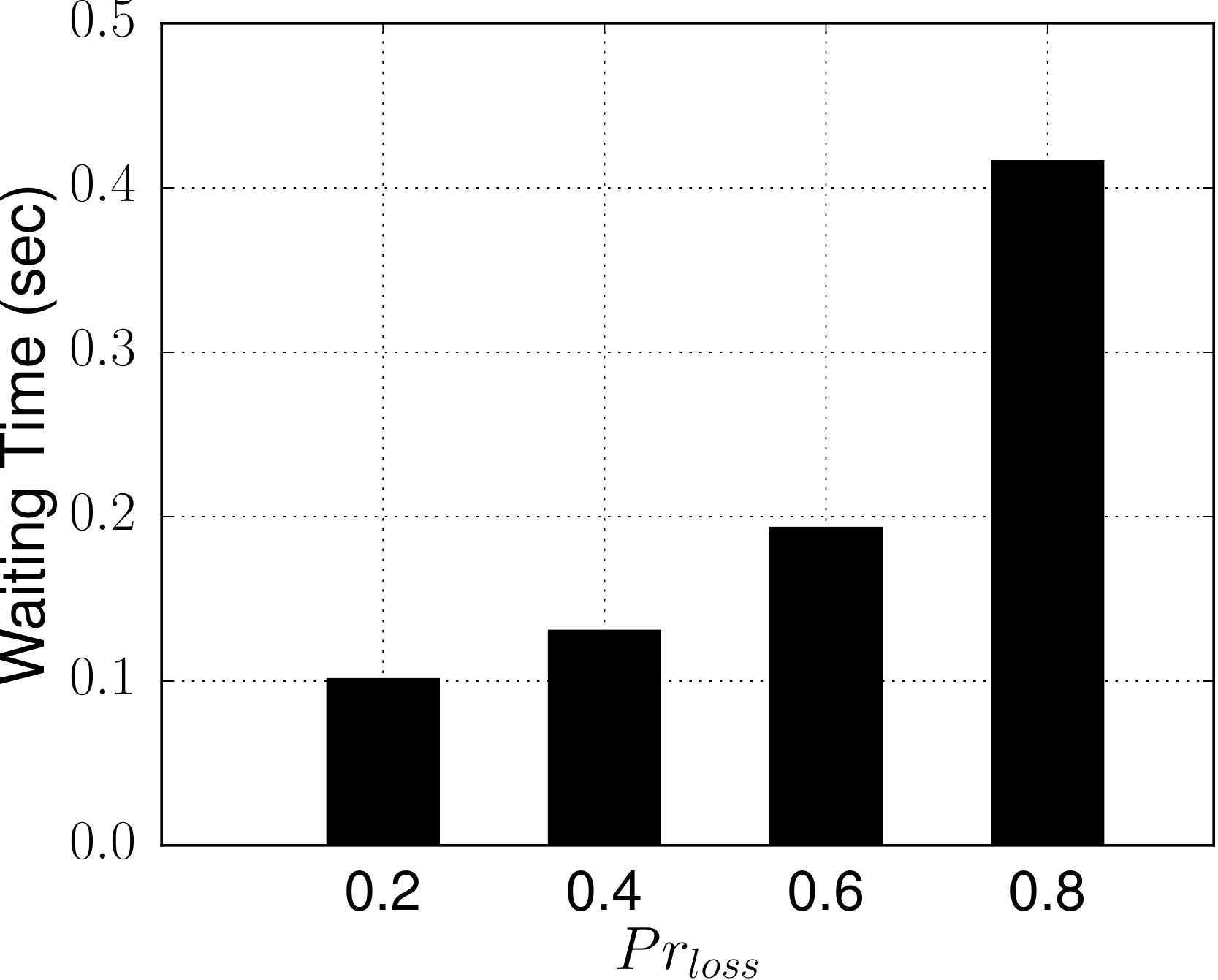}%
		\caption{}%
		\label{subfig_mobile_adv_num2_waiting}%
	\end{subfigure}\hspace{1mm}
	\begin{subfigure}{.24\textwidth}
		\includegraphics[width=\columnwidth]{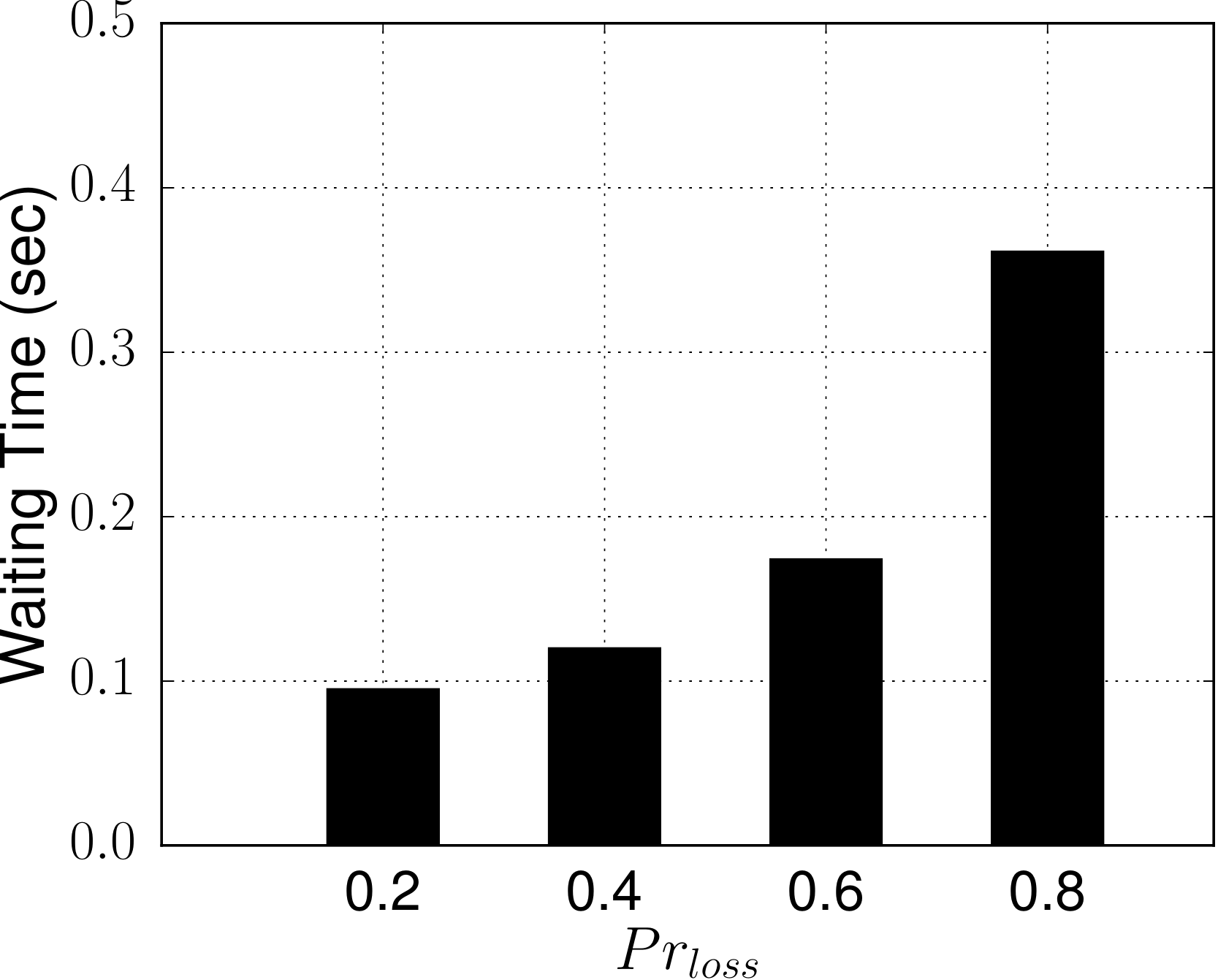}%
		\caption{}%
		\label{subfig_mobile_adv_num4_waiting}%
	\end{subfigure}\hspace{1mm}
	\begin{subfigure}{.24\textwidth}
		\includegraphics[width=\columnwidth]{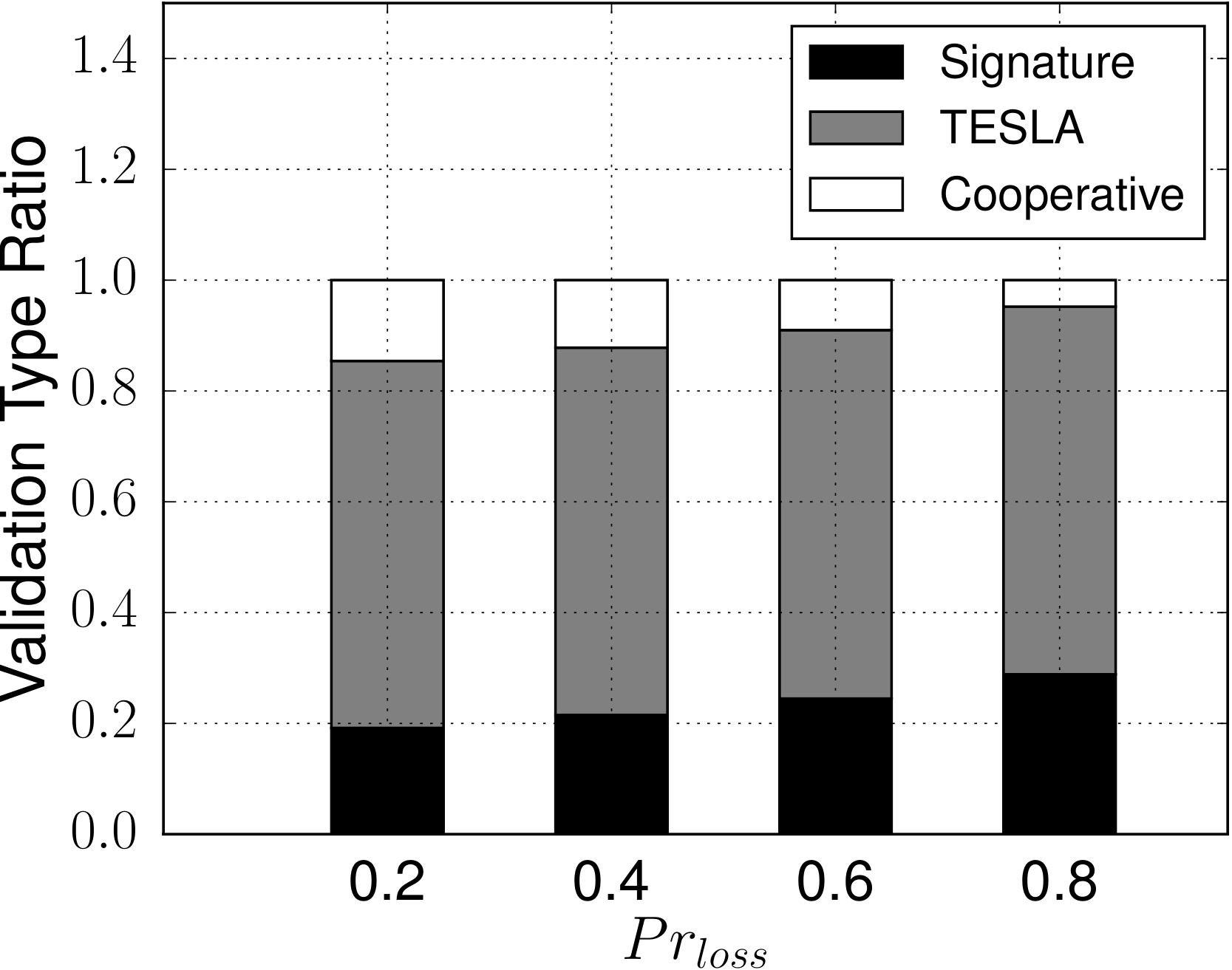}%
		\caption{}%
		\label{subfig_mobile_adv_num2_type}%
	\end{subfigure}\hspace{1mm}
	\begin{subfigure}{.24\textwidth}
		\includegraphics[width=\columnwidth]{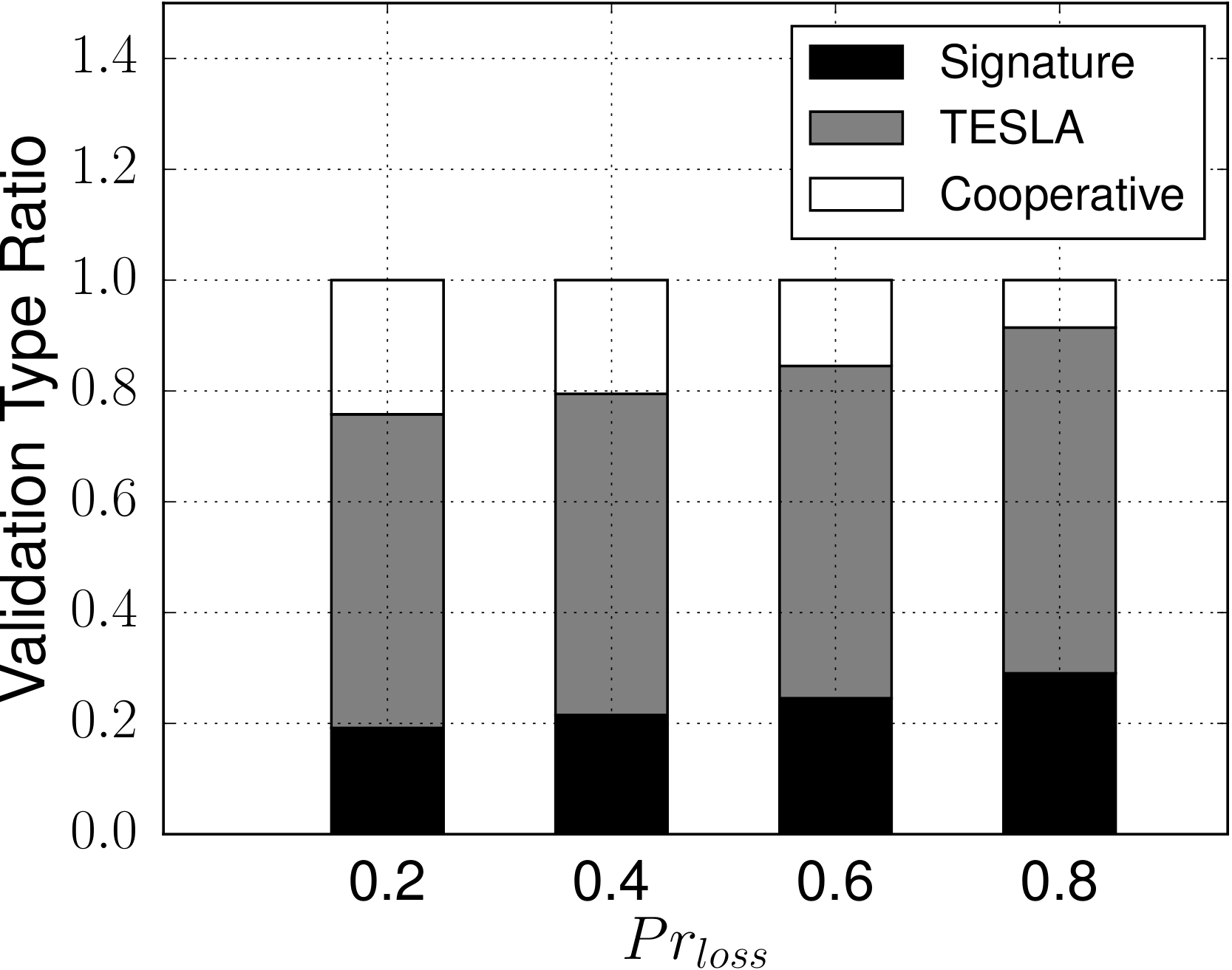}%
		\caption{}%
		\label{subfig_mobile_adv_num4_type}%
	\end{subfigure}
	\caption{Adversarial and Highway: Average waiting time as a function of $Pr_{loss}$ with (\subref{subfig_mobile_adv_num2_waiting}) $\alpha = 2$ and (\subref{subfig_mobile_adv_num4_waiting}) $\alpha = 4$. Ratio of validated beacons based on different validation types as a function of $Pr_{loss}$ with (\subref{subfig_mobile_adv_num2_type}) $\alpha = 2$ and (\subref{subfig_mobile_adv_num4_type}) $\alpha = 4$.}
	\label{fig_mobile_adv_waiting}
\end{figure*}

\subsection{Highway scenario}

\cref{fig_mobile} shows the simulated area for the highway scenario. We consider a six-lane $1.5\ km$ highway section, with vehicles entering from both ends. We set the vehicle arrival rate to make the average neighbor density roughly 60 ($N = 58.57$ on average exactly, measured from the simulation results). For this scenario, the first $1\ min$ of each simulation run is used as a warm-up phase to make nodes spatially distributed across the simulated area. Nodes start beaconing from $60\ s$ and the results between $60\ s$ -- $120\ s$ is used. Waiting times are collected from nodes when passing the gray area (\cref{fig_mobile}).

\textbf{Benign network:} In the highway scenario, nodes could continuously encounter new nodes during their trips. It is important that a node gains awareness of any new node entering its communication range on a timely manner. We first evaluate the average pseudonym validation delay, which reflects the speed of discovering new \acp{PC} (thus their owners). \emph{Pseudonym validation delay} is defined as the passed time from a new node (\ac{PC}) is encountered (in terms of received beacons) until at least one beacon from that node is verified. As described in Sec.~\ref{sec:scheme}, after a beacon is verified with signature, then the successive beacons from that node can be validated at least based on (relatively cheaper) TESLA MACs.

We record pseudonym validation delays for each pair of sender/receiver satisfying the condition: the distance between two nodes had been less than $150\ m$. This essentially eliminates node pairs that only had been at the border of communication range of each other. From \cref{subfig_mobile_num_psnym,subfig_mobile_loss_psnym}, we see pseudonym validation delays are almost always less than $1\ s$ with our scheme. This indicates that any pair of nodes that were closer than $150\ m$ can discover each other within $1\ s$. However, with only TESLA-based validation ($\alpha=0$), pseudonym validation delay significantly increases: only around $80\ \%$ of nodes can be discovered within $1\ s$.

\cref{subfig_mobile_num_waiting,subfig_mobile_loss_waiting} show average waiting time as a function of $\alpha$ and $Pr_{loss}$ respectively, and \cref{subfig_mobile_num_type,subfig_mobile_loss_type} show ratios of beacon validation types as a function of $\alpha$ and $Pr_{loss}$ respectively. We see the average waiting times and ratios of validation types are roughly same as those in the static scenario with $N=60$ (see \cref{fig_static_waiting}).

\textbf{Adversarial network:} We continue with evaluation under the adversarial network scenarios. Here, we first evaluate pseudonym validation ratio. Pseudonym validation ratio is defined as the ratio of validated \acp{PC} (i.e., discovered nodes) over total received legitimate \acp{PC} (i.e., encountered nodes) at the end of nodes' trips, reflecting the degree of awareness of neighboring nodes. \cref{fig_mobile_adv_ratio} shows pseudonym validation ratio with $\alpha = 0, 2, 4$ respectively. Without the cooperative verification scheme (\cref{subfig_mobile_adv_num_0_ratio} when $\alpha=0$), only slightly more than $50\ \%$ of \acp{PC} can be validated, because the majority of computation resource was dedicated for verifying fake signatures. Moreover, even those $50\ \%$ of \acp{PC} are validated with high delays (see \cref{subfig_mobile_adv_num_0_psnym}): only around $80\ \%$ (among the above-mentioned  $50\ \%$) of \acp{PC} can be validated within $5\ s$. 

\cref{subfig_mobile_adv_num_2_ratio,subfig_mobile_adv_num_4_ratio} show that, with our scheme, the nodes can effectively discover (in fact, almost all) valid \acp{PC} even under the DDoS attack. We see the pseudonym validation ratios are almost $100\ \%$, but slightly less than $100\ \%$ with high $Pr_{loss}$ values. This is because pseudonym validation delays are higher with higher $Pr_{loss}$ values (see \cref{subfig_mobile_adv_num_2_psnym,subfig_mobile_adv_num_4_psnym}): newly received \acp{PC} have not been validated at the end of node trips, while we can expect that these \acp{PC} could have been validated if the node trips continued further.

\cref{fig_mobile_adv_waiting} shows average waiting times and ratios of validation types for the highway scenario under the DDoS attack. We see our scheme still provides reasonable waiting times (\cref{subfig_mobile_adv_num2_waiting,subfig_mobile_adv_num4_waiting}) while more beacons have to be validated based on TESLA MACs and piggybacked verification results (\cref{subfig_mobile_adv_num2_type,subfig_mobile_adv_num4_type}).

}
\section{Discussions and conclusions}
\label{sec:conclusion}

{\color{blue}

In this paper, we consider a beacon rate of $10\ Hz$. However, adaptive beacon rate schemes~\cite{sommer2011traffic,nguyen2017mobility,schmidt2010exploration} adapt beaconing rates (generally lower than $10\ Hz$) based on the context (e.g., speed, vehicle density) for increased reliability and lower computation and communication overhead. Our scheme can be readily combined with adaptive beacon rate schemes. TESLA key chain generation and usage can be kept unchanged, but when beacon interval is larger than $0.1\ s$, the TESLA keys in between can be simply skipped. Skipped time slots are equivalent to lost beacons, thus transparent to the receiver (i.e., no change is needed for the beacon reception and validation processes), but the sender is taking the initiative.

We mandate each beacon to be signed by its sender so that the receiver can always choose to verify the signature if necessary, notably to achieve non-repudiation. Future work will address this explicitly, i.e., the conditions a receiver needs to verify signatures in order to minimize security risks incurred by internal adversaries (e.g., sharing false verification results or attaching fake signatures with correct TESLA keys and MACs).

}

We demonstrated how our cooperative beacon verification scheme could enable secure \ac{VC} at network densities even double compared to those prior approaches could be workable for. Even under DoS attacks, vehicles could still maintain a low waiting time for each received beacon so that the vehicles can gain awareness of neighboring vehicles within short time period. In addition, our scheme is orthogonal to all prior optimizations and could complement them.

\section*{References}

\bibliography{references}

\end{document}